# MBUR Generated Exponential Distribution: A New Generalization of Exponential Distribution

Iman M. Attia *

imanattiathesis1972@gmail.com ,imanattia1972@gmail.com

**Department of Mathematical Statistics, Faculty of Graduate Studies for Statistical Research, Cairo University, Egypt*

***Abstract:*** The unit distribution can serve as a foundation for adding a new parameter to continuous distributions and generating a generalized form with new properties. In this paper, the Median-Based Unit Rayleigh Distribution (MBUR) is used as a generator to create a generalized forms of the exponential distribution. The new MBUR-generated exponential distribution has a well-closed cumulative distribution function and quantile function. The PDF depicts a severe form of right-sided skewness with a long right tail. The moments are not expressed in a closed form and require the special function, the gamma function. The Hazard rates exhibit an initial increasing or decreasing pattern then a constant profile. The new distribution inherits the initial pattern from the MBUR that could be increasing or decreasing and inherits the latter constant profile from the exponential distribution. The limitation of this new distribution is that it does not show a bathtub hazard rate. The author discussed real data and deployed a distributional fit for this new MBUR-generated exponential distribution. The author also adopted the T-X{Y} procedure to create the MBUR exponential {Kumaraswamy} and MBUR exponential {Topp Loene} distributions.

.

## Introduction

Exponential distribution performs a fundamental task in lifetime or survival data studies. This is reasonably for its imperative lack of memory property, its constant hazard rates, and its analytical tractability. The broader distributions like the gamma, Weibull and Gombertz distributions are customarily applicable, in situations for which the one parameter family of exponential distribution is not satisfactory.

Numerous schemes have been used to add parameters to the exponential distribution like applying the Marshall Olkin generator. Marshall and Olkin (Marshall and Olkin 1997), defined the following one parameter family of survival functions

$$\bar{G}(x;\alpha) = \frac{\alpha\bar{F}(x)}{1-\bar{\alpha}\,\bar{F}(x)} \qquad \infty < x < \infty \;\; \& \;\; 0 < \alpha < \infty$$

Hence using the $\bar{F}(x) = e^{-\lambda x}$, the survival function of the exponential distribution, a new parameter is added to the exponential distribution. (Cordeiro, Lemonte, and Ortega 2014) discussed the mathematical properties and the new models of this Marshall Olkin family of distributions.

Moreover, (Gupta and Kundu 1999) was the first to put forward a generalization of the exponential called exponentiated exponential using the CDF, $F(x) = (1-e^{-\lambda x})^{\alpha}$ $x > 0$ , $\lambda, \alpha > 0$ .This is merely the $\alpha^{th}$ power of exponential distribution. see also (Gupta and Kundu 2001a),(Gupta and Kundu 2001b),(Gupta and Kundu 2003),(Gupta and Kundu 2004),(Gupta and Kundu 2007).

(Nadarajah and Kotz 2006) introduced the beta exponential distribution. They used the Beta generated family and the CDF of the exponential distribution to get the pdf of the beta exponential distribution:

$$f(x) = \frac{\lambda}{\mathrm{B}(a,b)} e^{-b\lambda x} \left(1-e^{-\lambda x}\right)^{a-1}; \;\; x, \lambda, a, b > 0$$

(Barreto-Souza, Santos, and Cordeiro 2008) introduced the beta generalized exponential distribution using the concept of the Beta generated family introduced by (Eugene, Lee, and Famoye 2002) and substituted the generalized exponentiated distribution introduced by (Gupta and Kundu 1999)

(Mirhossaini and Dolati 2008) introduced a new generalization of the exponential distribution using the mixture exponential with the scale parameter theta and shape parameter alpha with the following pdf

$$g(x;\alpha,\theta) = \theta e^{-\theta x}\left(1+\alpha\left[2e^{-\theta x}-1\right]\right); \qquad x, \theta > 0, \qquad -1 \le \alpha \le 1$$

(Cordeiro, Ortega, and Cunha 2013) advocated a new class of distributions that expands the exponentiated type distributions applying the double transformations incorporating both Lehmann type I and Lehmann type II. They defined exponentiated generalized (EG) class of the distribution by defining the following:

$$F(x) = \left[1-\left(1-G(x)\right)^{\alpha}\right]^{\beta}, \alpha > 0 \;\; \& \;\; \beta > 0$$

(Ristić and Kundu 2015) implemented the Marshall-Olkin method to obtain the two-parameters generalized exponential distribution with the following CDF:

$$G(x;\lambda,\alpha,\theta) = \frac{\left(1-e^{-\lambda x}\right)^{\alpha}}{\theta+(1-\theta)(1-e^{-\lambda x})^{\alpha}} \;, x; \lambda, \alpha, \theta > 0$$

So the two additional parameters are the $\alpha$ and the $\beta$ . The parent distribution G(x) can be the CDF of the exponential distribution and hence exponentiated generalized exponential is obtained.

The literatures include many papers about generalizing the exponential distribution. To mention some of them: (Merovci 2013) introduced the transmuted exponentiated exponential distribution. (Iqbal et al. 2014) introduced the generalized exponentiated moment exponential distribution. (Elbatal, Diab, and Abdul Alim 2013) discussed the transmuted generalized linear exponential distribution. (Tian, Tian, and Zhu 2014) explained the transmuted linear exponential distribution. (Khan, King, and Hudson 2015) discussed the transmuted generalized exponential distribution.

This paper is structured as follows: In section 1: the author explains the MBUR generated exponential distribution. In section 2 the author discusses the MBUR Exponential {Kumaraswamy} distribution. In section 3, the author elaborates on the MBUR exponential {Topp Leone} distribution. In section 4, the author illustrates real data analysis. In section 5, the author comprehends the conclusions and the future work.

# Section 1: MBUR Generated Exponential Distribution

### 1.1 Methodology and Derivation of the Distribution

Adopting the prototype of family generated distribution, the author used the exponential distribution as the parent distribution and the MBUR as the generator.

The exponential distribution has the PDF, CDF, and quantile function

$$f_X(x) = \lambda e^{-\lambda} \ , \qquad x > 0 \ \& \ \ \lambda > 0$$

$$F_X(x) = 1 - e^{-\lambda x}$$

$$Q(u;\lambda) = \frac{-1}{\lambda}\ln(1-u)$$

The Median Based Unit Rayleigh (MBUR) has the following PDF, CDF, and quantile function

$$f_Y(y) = \frac{6}{\alpha^2}(y)^{\frac{2}{\alpha^2}-1}\left(1-(y)^{\frac{1}{\alpha^2}}\right), \qquad 0 < \ y < 1 \ \ \& \ \alpha > 0$$

$$F_Y(y) = 3(y)^{\frac{2}{\alpha^2}} - 2(y)^{\frac{3}{\alpha^2}}$$

$$y = \left\{-.5\left(cos\left[\frac{cos^{-1}(1-2u)}{3}\right] - \sqrt{3}\,sin\left[\frac{cos^{-1}(1-2u)}{3}\right]\right) + .5\right\}^{\alpha^2}$$

Now the MBUR generated exponential has the CDF, and the PDF

$$G_{X_g}(x) = \int_0^{F(x)} f_{MBUR}(t)\,dt = \ F_{MBUR}\big(F_X(x)\big) = 3\big(F_X(x)\big)^{\frac{2}{\alpha^2}} - 2\big(F_X(x)\big)^{\frac{3}{\alpha^2}}$$

$$G_{X_g}(x) = 3\big(1-e^{-\lambda x}\big)^{\frac{2}{\alpha^2}} - 2\big(1-e^{-\lambda x}\big)^{\frac{3}{\alpha^2}}$$

$$g_{X_g}(x) = \ f_{MBUR}\big(F_X(x)\big)\,f_X(x)$$

$$g_{X_g}(x) = \frac{6}{\alpha^2}\left(F_X(x)\right)^{\frac{2}{\alpha^2}-1}\left(1-\left(F_X(x)\right)^{\frac{1}{\alpha^2}}\right)\lambda e^{-\lambda x}$$

$$g_{X_g}(x) = \frac{6}{\alpha^2}\left(1-e^{-\lambda x}\right)^{\frac{2}{\alpha^2}-1}\left(1-\left(1-e^{-\lambda x}\right)^{\frac{1}{\alpha^2}}\right)\lambda e^{-\lambda x}$$

### 1.2 properties of the distribution

#### 1.2.1 The hazard function

**Proposition:** the hazard function is defined as $hr = \frac{g_{X_g}(x)}{1-G_{X_g}(x)}$

$$hr = \frac{g_{X_g}(x)}{1-G_{X_g}(x)} = \frac{\frac{6}{\alpha^2}\left(1-e^{-\lambda}\ \right)^{\frac{2}{\alpha^2}-1}\left(1-\left(1-e^{-\lambda x}\right)^{\frac{1}{\alpha^2}}\right)\lambda e^{-\lambda}}{1-3\left(1-e^{-\lambda}\ \right)^{\frac{2}{\alpha^2}}+2\left(1-e^{-\lambda x}\right)^{\frac{3}{\alpha^2}}}$$

#### 1.2.2 The quantile function

**Theorem:** The quantile function is defined as

$$Q(u;\lambda,\alpha) = \frac{-1}{\lambda}\ln\left(1-\left\{-.5\left(cos\left[\frac{cos^{-1}(1-2u)}{3}\right]-\sqrt{3}\,sin\left[\frac{cos^{-1}(1-2u)}{3}\right]\right)+.5\right\}^{\alpha^2}\right)$$

**Proof:**

$$u = F_{MBUR}\left(F_X(x)\right)$$

Take the inverse on both sides $F_{MBUR}^{-1}(u) = F_{MBUR}^{-1}\left[F_{MBUR}\left(F_X(x)\right)\right]$, then take the inverse for the second time $F_X^{-1}[F_{MBUR}^{-1}(u)] = F_X^{-1}[F_X(x)]$, this leads to

$x = F_X^{-1}[F_{MBUR}^{-1}(u)]$.

**Step 1**: get the inverse of MBUR

$$y = \left\{-.5\left(cos\left[\frac{cos^{-1}(1-2u)}{3}\right]-\sqrt{3}\,sin\left[\frac{cos^{-1}(1-2u)}{3}\right]\right)+.5\right\}^{\alpha^2}$$

**Step 2**: substitute this inverse into the quantile function of the exponential

$$Q(u;\lambda) = \frac{-1}{\lambda}\ln\left(1-\left\{-.5\left(cos\left[\frac{cos^{-1}(1-2u)}{3}\right]-\sqrt{3}\,sin\left[\frac{cos^{-1}(1-2u)}{3}\right]\right)+.5\right\}^{\alpha^2}\right)$$

#### 1.2.3 The raw moments

**Theorem**: the $r^{th}$ raw moment is defined as

$$E(x^r) = \frac{6\Gamma(r+1)}{\alpha^2\lambda^r}\ \left\{\frac{w_k}{(1+k)^{(r+1)}}-\frac{w_{k,m}}{(1+k+m)^{(r+1)}}\right\}$$

**Proof:**

By definition the raw moments are defined as $E(x^r) = \int_0^\infty x^r\, g_{X_g}(x)\, dx$

$$E(x^r) = \frac{6\lambda}{\alpha^2} \int_0^\infty x^r\, e^{-\lambda} \left(1 - e^{-\lambda x}\right)^{\frac{2}{\alpha^2}-1} \left(1 - \left(1 - e^{-\lambda x}\right)^{\frac{1}{\alpha^2}}\right) dx$$

Using the generalized binomial expansion of the following terms:

First term: $\left(1 - e^{-\lambda x}\right)^{\frac{2}{\alpha^2}-1} = \sum_{k=0}^{\infty}(-1)^k \binom{\frac{2}{\alpha^2}-1}{k} e^{-\lambda x k}$

Second term: $\left(1 - \left(1 - e^{-\lambda x}\right)^{\frac{1}{\alpha^2}}\right) = 1 - \sum_{m=0}^{\infty}(-1)^m \binom{\frac{1}{\alpha^2}}{m} e^{-\lambda x}$

$$E(x^r) = \frac{6\lambda}{\alpha^2} \int_0^\infty x^r\, e^{-\lambda x} \sum_{k=0}^{\infty}(-1)^k \binom{\frac{2}{\alpha^2}-1}{k} e^{-\lambda x k} \left\{1 - \sum_{m=0}^{\infty}(-1)^m \binom{\frac{1}{\alpha^2}}{m} e^{-\lambda x m}\right\} dx$$

$$E(x^r) = \frac{6\lambda}{\alpha^2} \int_0^\infty x^r\, e^{-\lambda x} \sum_{k=0}^{\infty}(-1)^k \binom{\frac{2}{\alpha^2}-1}{k} e^{-\lambda x k}\, dx$$

$$-\frac{6\lambda}{\alpha^2} \int_0^\infty x^r\, e^{-\lambda x} \sum_{k=0}^{\infty}(-1)^k \binom{\frac{2}{\alpha^2}-1}{k} e^{-\lambda x} \sum_{m=0}^{\infty}(-1)^m \binom{\frac{1}{\alpha^2}}{m} e^{-\lambda x m}\, dx$$

$$E(x^r) = \frac{6\lambda}{\alpha^2} \sum_{k=0}^{\infty}(-1)^k \binom{\frac{2}{\alpha^2}-1}{k} \int_0^\infty x^r\, e^{-\lambda x(1+k)}\, dx$$

$$-\frac{6\lambda}{\alpha^2} \sum_{k=0}^{\infty}\sum_{m=0}^{\infty}(-1)^{k+m} \binom{\frac{2}{\alpha^2}-1}{k} \binom{\frac{1}{\alpha^2}}{m} \int_0^\infty x^r\, e^{-\lambda x(1+k+m)}\, dx$$

$$E(x^r) = \frac{6\lambda}{\alpha^2} \left\{\frac{w_k \Gamma(r+1)}{[\lambda(1+k)]^{(r+1)}} - \frac{w_{k,m} \Gamma(r+1)}{[\lambda(1+k+m)]^{(r+1)}}\right\}$$

$$E(x^r) = \frac{6\lambda}{\alpha^2} \frac{\Gamma(r+1)}{\lambda^{(r+1)}} \left\{\frac{w_k}{[(1+k)]^{(r+1)}} - \frac{w_{k,m}}{[(1+k+m)]^{(r+1)}}\right\}$$

$$E(x^r) = \frac{6\Gamma(r+1)}{\alpha^2 \lambda^r} \left\{\frac{w_k}{(1+k)^{(r+1)}} - \frac{w_{k,m}}{(1+k+m)^{(r+1)}}\right\}$$

Where

$w_k = \sum_{k=0}^{\infty}(-1)^k \binom{\frac{2}{\alpha^2}-1}{k}$ and $w_{k,m} = \sum_{k=0}^{\infty}\sum_{m=0}^{\infty}(-1)^{k+m} \binom{\frac{2}{\alpha^2}-1}{k} \binom{\frac{1}{\alpha^2}}{m}$

Figure (1-6) show PDF, CDF, and hazard function of the MBUR generated exponential. The distribution has skewed PDF and decreasing PDF shapes. the hazard rate function has an initial increasing or decreasing part followed by a constant part

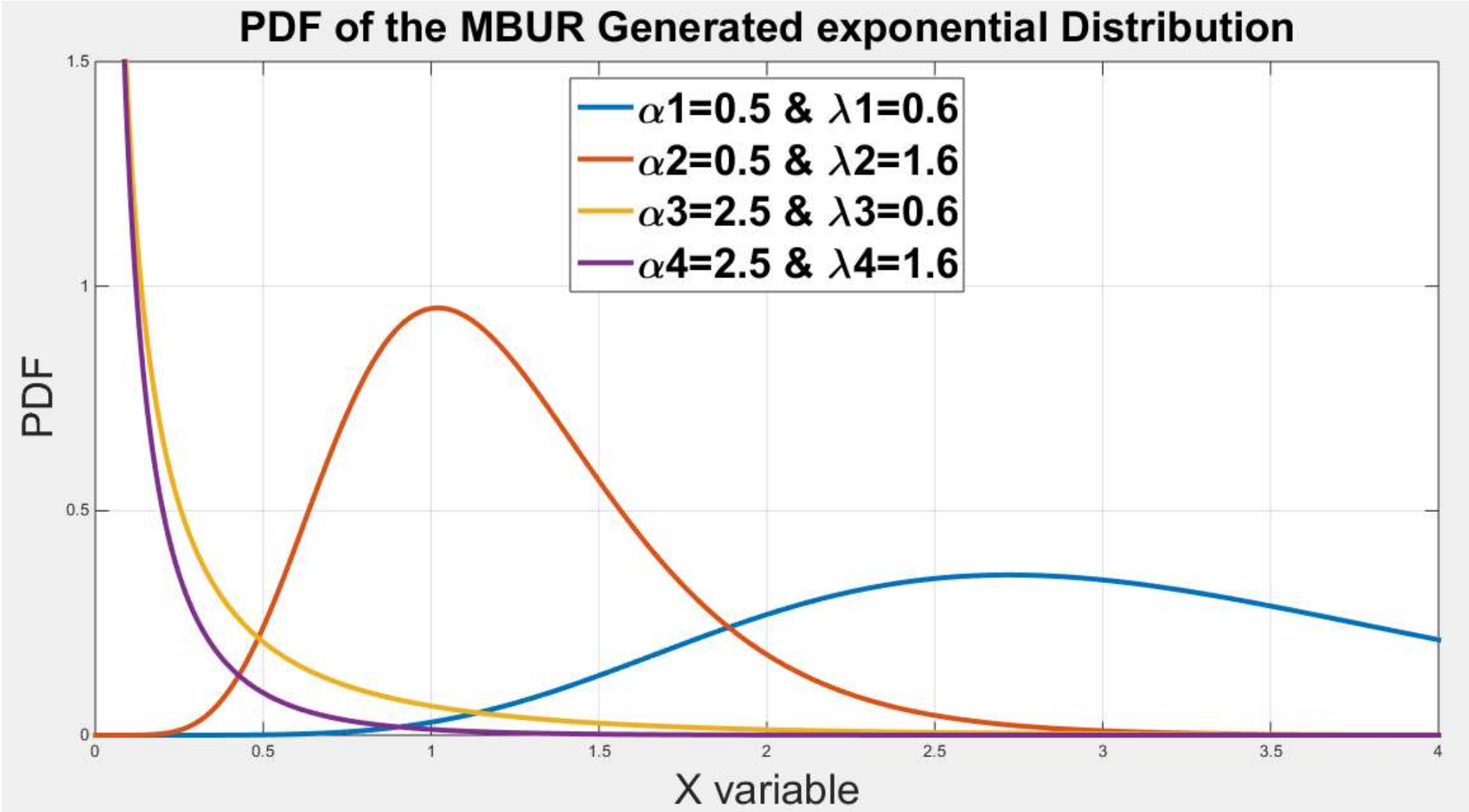

Fig. 1 shows the pdf of the MBUR generated exponential distribution

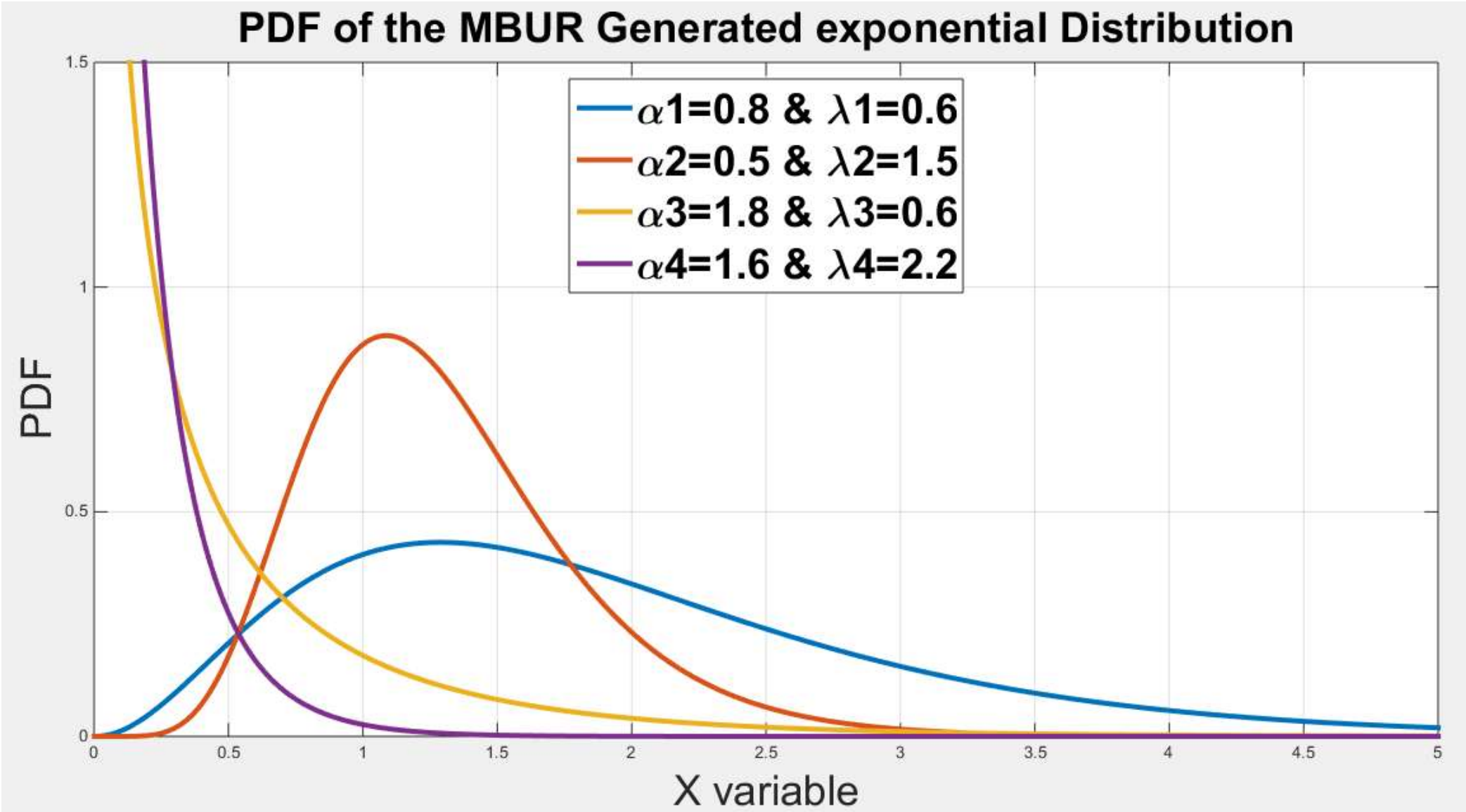

Fig. 2 shows the pdf of the MBUR generated exponential distribution.

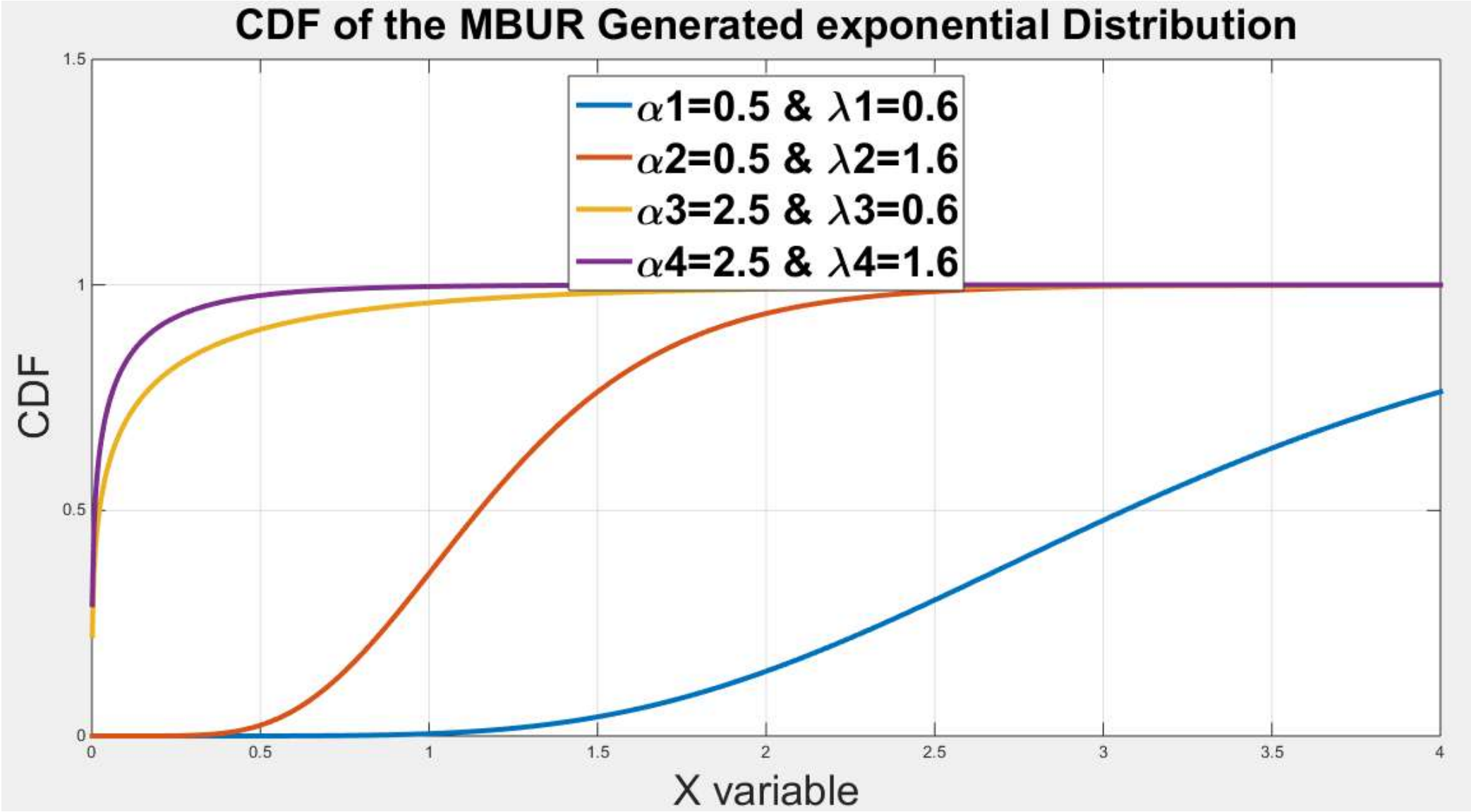

Fig. 3 shows the CDF of the MBUR generated exponential distribution.

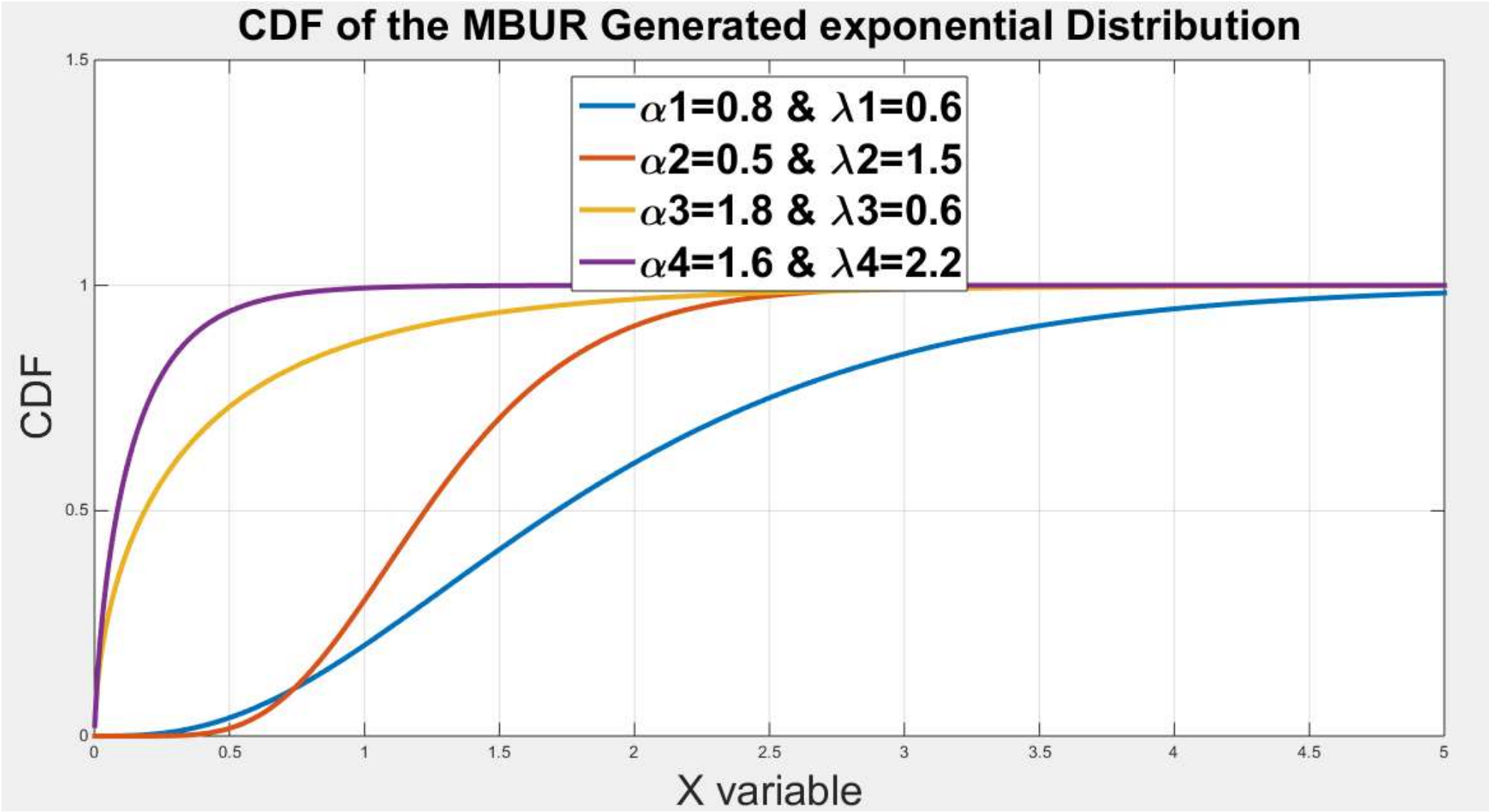

Fig. 4 shows the CDF of the MBUR generated exponential distribution.

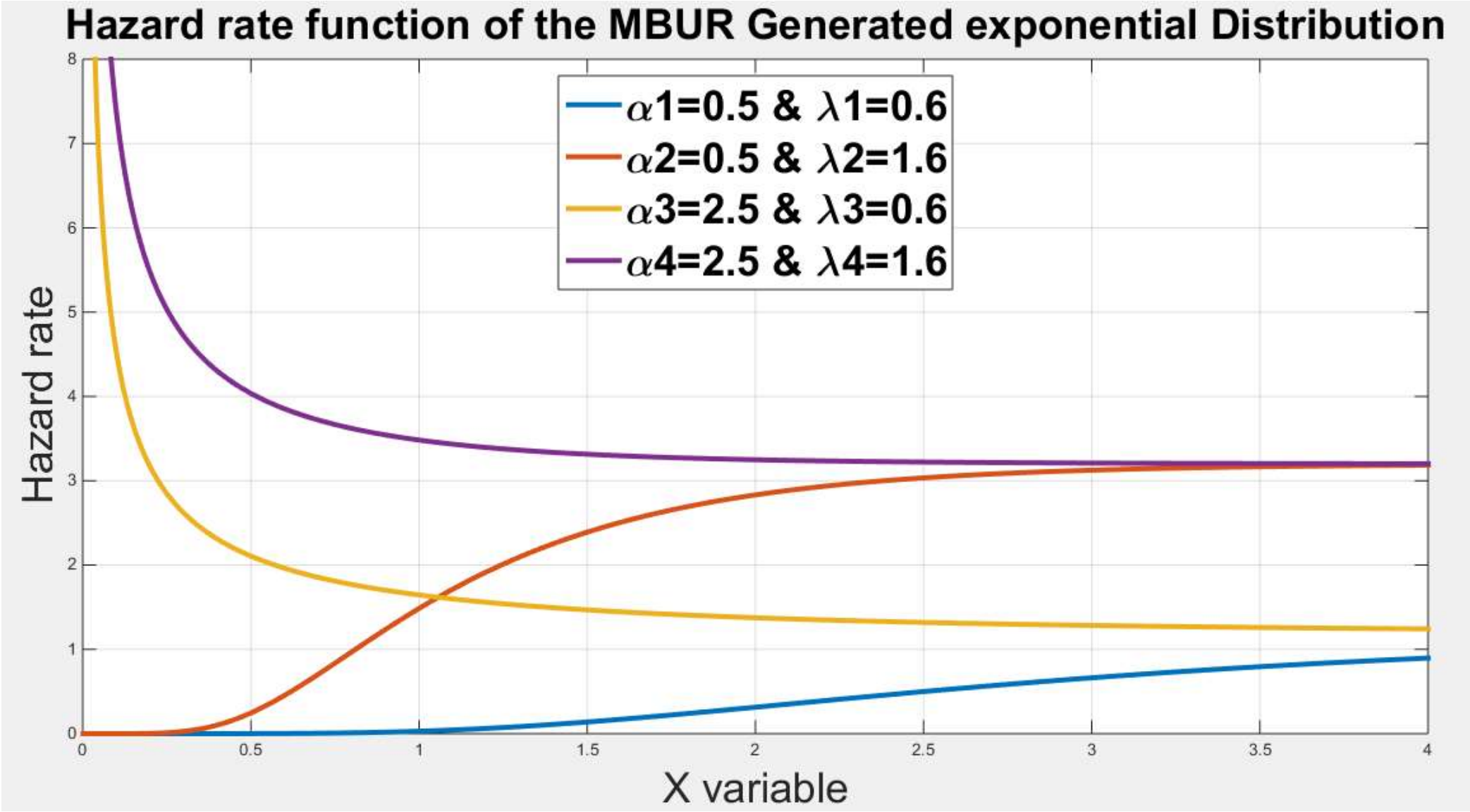

Fig. 5 shows the hazard rate function of the MBUR generated exponential distribution.

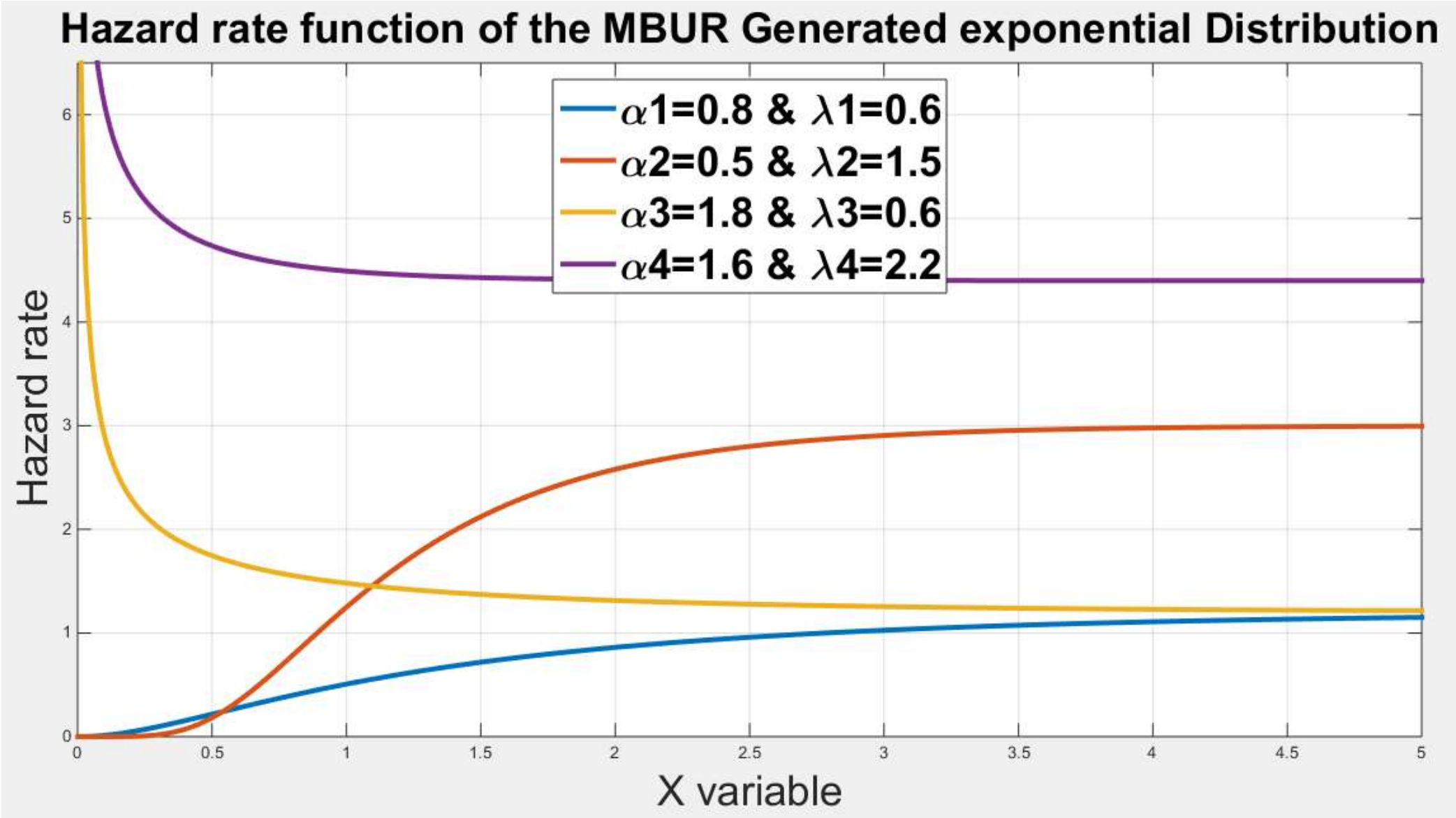

Fig. 6 shows the hazard rate function of the MBUR generated exponential distribution

# Section 2: MBUR-Exponential {Kumaraswamy}

## 2.1 Methodology and Derivation of the Distribution

Using the T-X {Y} technique, introduced by (Aljarrah, Lee, and Famoye 2014), to generate the MBUR- exponential {Kumaraswamy} where:

The generator (T) is the MBUR dsitribution. It has the following PDF, CDF, and quantile functions respectively:

$$f_T(t) = \frac{6}{\alpha^2}(t)^{\frac{2}{\alpha^2}-1}\left(1-(t)^{\frac{1}{\alpha^2}}\right), \qquad 0 < \ t < 1 \ \ \& \, \alpha > 0$$

$$F_T(t) = 3(t)^{\frac{2}{\alpha^2}} - 2(t)^{\frac{3}{\alpha^2}}$$

$$t = \left\{-.5\left(cos\left[\frac{cos^{-1}(1-2u)}{3}\right] - \sqrt{3}\,sin\left[\frac{cos^{-1}(1-2u)}{3}\right]\right) + .5\right\}^{\alpha^2}$$

The quantile link distribution (Y) is the Kumaraswamy. It has the following the PDF, CDF, and quantile function respectively:

$$f_Y(y) = aby^{a-1}(1-y^a)^{b-1}\ , \qquad 0 < y < 1\ , \qquad a > 0\ \&\ b > 0$$

$$f_Y(y) = 1 - (1-y^a)^b$$

$$y = Q(u; a, b) = \left[1 - (1-u)^{1/b}\right]^{1/a}$$

The baseline distribution (X) is the Exponential with PDF, CDF, and quantile functions respectively as:

$$f_X(x) = \theta e^{-\theta}\ , \qquad x > 0\ \&\ \ \theta > 0$$

$$F_X(x) = \left[1 - e^{-\theta x}\right]$$

$$x = Q(u; \theta) = \frac{-1}{\theta}\ln(1-u)$$

The steps of creating the new distribution are as the following:

**Step1**: Substitute the CDF of the exponential into the quantile of the Kumaraswamy

$$T = Q_Y(F_X(x)) = \left[1-(1-u)^{1/b}\right]^{1/a} = \left[1-\left(1-\left[1-e^{-\theta x}\right]\right)^{1/b}\right]^{1/a} = \left[1-e^{-\frac{\theta x}{b}}\right]^{1/a}$$

**Step 2**: Substitute this T into pdf of the MBUR

$$f_T(t) = \frac{6}{\alpha^2}\left(\left[1-e^{-\frac{\theta x}{b}}\right]^{\frac{1}{a}}\right)^{\frac{2}{\alpha^2}-1}\left(1-\left(\left[1-e^{-\frac{\theta x}{b}}\right]^{\frac{1}{a}}\right)^{\frac{1}{\alpha^2}}\right)$$

**Step 3**: Get the derivative of the of the quantile function which is the reciprocal of the PDF of the quantile distribution then substitute into the equation of the g(x)

$$g_{X_g}(x) = f_X(x)\ \frac{r\left(Q_Y(F_X(x))\right)}{f_Y\left(Q_Y(F_X(x))\right)} = f_X(x)\ \frac{f_T\left(Q_Y(F_X(x))\right)}{f_Y\left(Q_Y(F_X(x))\right)}$$

$$g_{X_g}(x) = \frac{\frac{6}{\alpha^2}\left(\left[1-e^{-\frac{\theta x}{b}}\right]^{\frac{1}{a}}\right)^{\frac{2}{\alpha^2}-1}\left(1-\left(\left[1-e^{-\frac{\theta x}{b}}\right]^{\frac{1}{a}}\right)^{\frac{1}{\alpha^2}}\right)\theta e^{-\theta}}{ab\left(\left[1-e^{-\frac{\theta x}{b}}\right]^{\frac{1}{a}}\right)^{a-1}\left(1-\left(\left[1-e^{-\frac{\theta x}{b}}\right]^{\frac{1}{a}}\right)^{a}\right)^{b-1}}$$

$$g_{X_g}(x) = \frac{\frac{6}{\alpha^2}\left(\left[1-e^{-\frac{\theta x}{b}}\right]^{\frac{1}{a}}\right)^{\frac{2}{\alpha^2}-1}\left(1-\left(\left[1-e^{-\frac{\theta x}{b}}\right]^{\frac{1}{a}}\right)^{\frac{1}{\alpha^2}}\right)\theta e^{-\theta x}}{ab\left(\left[1-e^{-\frac{\theta x}{b}}\right]^{\frac{1}{a}}\right)^{a-1}\left(1-\left[1-e^{-\frac{\theta x}{b}}\right]\right)^{b-1}}$$

$$g_{X_g}(x) = \frac{\frac{6}{\alpha^2}\left(\left[1-e^{-\frac{\theta x}{b}}\right]^{\frac{1}{a}}\right)^{\frac{2}{\alpha^2}}\left(1-\left(\left[1-e^{-\frac{\theta x}{b}}\right]^{\frac{1}{a}}\right)^{\frac{1}{\alpha^2}}\right)\theta e^{-\theta}}{ab\left(\left[1-e^{-\frac{\theta x}{b}}\right]^{\frac{1}{a}}\right)^{a}\left(e^{-\frac{\theta x}{b}}\right)^{b-1}}$$

$$g_{X_g}(x) = \frac{\frac{6}{\alpha^2}\left(1-e^{-\frac{\theta x}{b}}\right)^{\frac{2}{\alpha^2 a}-1}\left(1-\left(\left[1-e^{-\frac{\theta x}{b}}\right]^{\frac{1}{a}}\right)^{\frac{1}{\alpha^2}}\right)\theta e^{-\theta x}}{ab(e)^{-\theta x+\frac{\theta x}{b}}}$$

$$g_{X_g}(x) = \frac{6}{\alpha^2 ab}\left(1-e^{-\frac{\theta x}{b}}\right)^{\frac{2}{\alpha^2 a}-1}\left(1-\left(\left[1-e^{-\frac{\theta x}{b}}\right]^{\frac{1}{a}}\right)^{\frac{1}{\alpha^2}}\right)\theta e^{-\frac{\theta x}{b}}$$

$$g_{X_g}(x) = \frac{6\theta}{\alpha^2 ab}\ e^{-\frac{\theta x}{b}}\left(1-e^{-\frac{\theta x}{b}}\right)^{\frac{2}{\alpha^2 a}-1}\left(1-\left[1-e^{-\frac{\theta x}{b}}\right]^{\frac{1}{\alpha^2 a}}\right)$$

$$G_{X_g}(x) = F_T(t) = 3\left(\left[1-e^{-\frac{\theta x}{b}}\right]^{1/a}\right)^{\frac{2}{\alpha^2}} - 2\left(\left[1-e^{-\frac{\theta x}{b}}\right]^{1/a}\right)^{\frac{3}{\alpha^2}}$$

$$G_{X_g}(x) = F_T(t) = 3\left(\left[1-e^{-\frac{\theta x}{b}}\right]^{\frac{1}{a\alpha^2}}\right)^2 - 2\left(\left[1-e^{-\frac{\theta x}{b}}\right]^{\frac{1}{a\alpha^2}}\right)^3$$

## 2.2 properties of the distribution

### 2.2.1 The hazard function

**Proposition:** The hazard function is defined as $hr = \frac{g_{X_g}(x)}{1-G_{X_g}(x)}$

$$hr = \frac{\frac{6\theta}{\alpha^2 ab} e^{-\frac{\theta x}{b}} \left(1-e^{-\frac{\theta x}{b}}\right)^{\frac{2}{\alpha^2 a}-1}\left(1-\left[1-e^{-\frac{\theta x}{b}}\right]^{\frac{1}{\alpha^2 a}}\right)}{1-3\left(\left[1-e^{-\frac{\theta x}{b}}\right]^{\frac{1}{a}}\right)^{\frac{2}{\alpha^2}}+2\left(\left[1-e^{-\frac{\theta x}{b}}\right]^{1/a}\right)^{\frac{3}{\alpha^2}}}$$

### 2.2.2 The quantile function

**Theorem:** The quantile function is defined as

$$\frac{-b}{\theta}\ln\left(1-\left\{-.5\left(cos\left[\frac{cos^{-1}(1-2u)}{3}\right]-\sqrt{3}\,sin\left[\frac{cos^{-1}(1-2u)}{3}\right]\right)+.5\right\}^{a\alpha^2}\right)$$

**proof:** By definition, CDF of the new distribution is defined as

$G_{X_g}(x) = R\left(Q_Y\left(F_X(x)\right)\right) = F_T\left(Q_Y\left(F_X(x)\right)\right)$ applying the following steps

**Step 1:** Take the inverse of both sides, use the inverse function of the MBUR

$$u = F_T\left(Q_Y\left(F_X(x)\right)\right)$$

$$F_T^{-1}(u) = F_T^{-1} F_T\left(Q_Y\left(F_X(x)\right)\right)$$

$$\left(Q_Y\left(F_X(x)\right)\right) = t = \left\{-.5\left(cos\left[\frac{cos^{-1}(1-2u)}{3}\right]-\sqrt{3}\,sin\left[\frac{cos^{-1}(1-2u)}{3}\right]\right)+.5\right\}^{\alpha^2}$$

**Step 2:** Take the CDF of both side: substitute this (t) into the CDF of the quantile distribution

$$F_T^{-1}(u) = \left(Q_Y\left(F_X(x)\right)\right)$$

$$F_Y[F_T^{-1}(u)] = F_Y\left(Q_Y\left(F_X(x)\right)\right)$$

$$F_Y[F_T^{-1}(u)] = \left(\left(F_X(x)\right)\right)$$

Where $F_Y(y) = 1-(1-y^a)^b$ *thus* $F_Y(t) = 1-(1-t^a)^b$ so

$$F_Y(y) = 1 - \left(1 - \left(\left\{-.5\left(cos\left[\frac{cos^{-1}(1-2u)}{3}\right] - \sqrt{3}\, sin\left[\frac{cos^{-1}(1-2u)}{3}\right]\right) + .5\right\}^{\alpha^2}\right)^{a}\right)^{b}$$

$$F_Y(y) = 1 - \left(1 - \left\{-.5\left(cos\left[\frac{cos^{-1}(1-2u)}{3}\right] - \sqrt{3}\, sin\left[\frac{cos^{-1}(1-2u)}{3}\right]\right) + .5\right\}^{a\alpha^2}\right)^{b}$$

**Step3:** Take the inverse once again on both side, i.e. substitute the CDF from the previous step into the quantile of the exponential

$$F_X^{-1}\{F_Y[F_T^{-1}(u)]\} = F_X^{-1}\left(\left(F_X(x)\right)\right)$$

$$Q(u) = x = F_X^{-1}\{F_Y[F_T^{-1}(u)]\}$$

$$Q(u;\theta,\alpha,a,b) = \frac{-1}{\theta}\ln(1-u)$$

$$Q(u) = \frac{-1}{\theta}\ln\left(1 - \left\{1 - \left(1 - \left\{-.5\left(cos\left[\frac{cos^{-1}(1-2u)}{3}\right] - \sqrt{3}\, sin\left[\frac{cos^{-1}(1-2u)}{3}\right]\right) + .5\right\}^{a\alpha^2}\right)^{b}\right\}\right)$$

$$Q(u) = \frac{-1}{\theta}\ln\left(1 - \left\{1 - \left(1 - \left\{-.5\left(cos\left[\frac{cos^{-1}(1-2u)}{3}\right] - \sqrt{3}\, sin\left[\frac{cos^{-1}(1-2u)}{3}\right]\right) + .5\right\}^{a\alpha^2}\right)^{b}\right\}\right)$$

$$Q(u) = \frac{-1}{\theta}\ln\left(\left\{1 - 1 + \left(1 - \left\{-.5\left(cos\left[\frac{cos^{-1}(1-2u)}{3}\right] - \sqrt{3}\, sin\left[\frac{cos^{-1}(1-2u)}{3}\right]\right) + .5\right\}^{a\alpha^2}\right)^{b}\right\}\right)$$

$$Q(u) = \frac{-1}{\theta}\ln\left(1 - \left\{-.5\left(cos\left[\frac{cos^{-1}(1-2u)}{3}\right] - \sqrt{3}\, sin\left[\frac{cos^{-1}(1-2u)}{3}\right]\right) + .5\right\}^{a\alpha^2}\right)^{b}$$

$$Q(u) = \frac{-b}{\theta}\ln\left(1 - \left\{-.5\left(cos\left[\frac{cos^{-1}(1-2u)}{3}\right] - \sqrt{3}\, sin\left[\frac{cos^{-1}(1-2u)}{3}\right]\right) + .5\right\}^{a\alpha^2}\right)$$

### 2.2.3 The raw moments

**Theorem:** the r^th raw moment can be expressed as:

$$E(x^r) = \frac{6\,\Gamma(r+1)}{\alpha^2 a}\left(\frac{b}{\theta}\right)^{r}\left\{\frac{w_k}{(1+k)^{(r+1)}} - \frac{w_{k,m}}{(1+k+m)^{(r+1)}}\right\}$$

**Proof:**

The PDF can be rewritten as linear series expansion.

Using the generalized binomial expansion of the following terms:

First term:

$$\left(1 - e^{-\frac{\theta x}{b}}\right)^{\frac{2}{\alpha^2 a} - 1} = \sum_{k=0}^{\infty}(-1)^k \binom{\frac{2}{\alpha^2 a} - 1}{k} e^{-\frac{\theta x k}{b}}$$

Second term:

$$\left[1-e^{-\frac{\theta x}{b}}\right]^{\frac{1}{\alpha^2 a}} = \sum_{m=0}^{\infty}(-1)^m\binom{\frac{1}{\alpha^2 a}}{m}e^{-\frac{\theta x m}{b}}$$

$$g_{X_g}(x) = \frac{6\theta}{\alpha^2 ab}e^{-\frac{\theta x}{b}}\sum_{k=0}^{\infty}(-1)^k\binom{\frac{2}{\alpha^2 a}-1}{k}e^{-\frac{\theta x k}{b}}\left\{1-\sum_{m=0}^{\infty}(-1)^m\binom{\frac{1}{\alpha^2 a}}{m}e^{-\frac{\theta x m}{b}}\right\}$$

$$g_{X_g}(x) = \frac{6\theta}{\alpha^2 ab}\sum_{k=0}^{\infty}(-1)^k\binom{\frac{2}{\alpha^2 a}-1}{k}e^{-\frac{\theta x(1+k)}{b}}$$

$$-\frac{6\theta}{\alpha^2 ab}\sum_{k=0}^{\infty}\sum_{m=0}^{\infty}(-1)^{k+m}\binom{\frac{2}{\alpha^2 a}-1}{k}\binom{\frac{1}{\alpha^2 a}}{m}e^{-\frac{\theta x(1+k+m)}{b}}$$

$$g_{X_g}(x) = \frac{6\theta}{\alpha^2 ab}\left\{w_k\, e^{-\frac{\theta x(1+k)}{b}} - w_{k,m}\, e^{-\frac{\theta x(1+k+m)}{b}}\right\}$$

Where:

$$w_k = \sum_{k=0}^{\infty}(-1)^k\binom{\frac{2}{\alpha^2 a}-1}{k}$$

$$w_{k,m} = \sum_{k=0}^{\infty}\sum_{m=0}^{\infty}(-1)^{k+m}\binom{\frac{2}{\alpha^2 a}-1}{k}\binom{\frac{1}{\alpha^2 a}}{m}$$

By definition : $E(x^r) = \int_0^{\infty} x^r\, g_{X_g}(x)\, dx$

$$E(x^r) = \frac{6\theta}{\alpha^2 ab}\int_0^{\infty} x^r\left\{w_k\, e^{-\frac{\theta x(1+k)}{b}} - w_{k,m}\, e^{-\frac{\theta x(1+k+m)}{b}}\right\} dx$$

$$E(x^r) = \frac{6\theta}{\alpha^2 ab}\int_0^{\infty} x^r\, w_k\, e^{-\frac{\theta x(1+k)}{b}}\, dx - \frac{6\theta}{\alpha^2 ab}\int_0^{\infty} x^r\, w_{k,m}\, e^{-\frac{\theta x(1+k+m)}{b}}\, dx$$

$$E(x^r) = \frac{6\theta}{\alpha^2 ab}\left\{\frac{w_k\Gamma(r+1)}{\left[\frac{\theta}{b}(1+k)\right]^{(r+1)}} - \frac{w_{k,m}\Gamma(r+1)}{\left[\frac{\theta}{b}(1+k+m)\right]^{(r+1)}}\right\}$$

$$E(x^r) = \frac{6\theta\Gamma(r+1)b^{(r+1)}}{\alpha^2 ab\,\theta^{(r+1)}}\left\{\frac{w_k}{[(1+k)]^{(r+1)}} - \frac{w_{k,m}}{[(1+k+m)]^{(r+1)}}\right\}$$

$$E(x^r) = \frac{6\Gamma(r+1)b^{(r)}}{\alpha^2 a\,\theta^{(r)}}\left\{\frac{w_k}{(1+k)^{(r+1)}} - \frac{w_{k,m}}{(1+k+m)^{(r+1)}}\right\}$$

$$E(x^r) = \frac{6\,\Gamma(r+1)}{\alpha^2 a}\left(\frac{b}{\theta}\right)^r \left\{\frac{w_k}{(1+k)^{(r+1)}} - \frac{w_{k,m}}{(1+k+m)^{(r+1)}}\right\}$$

Figures (7-12) show the PDF, CDF, and hazard rate function. The PDF has various skewed shaped presentation. The Hazard function takes the S shaped appearance, they are initially increased then attained a constant rate.

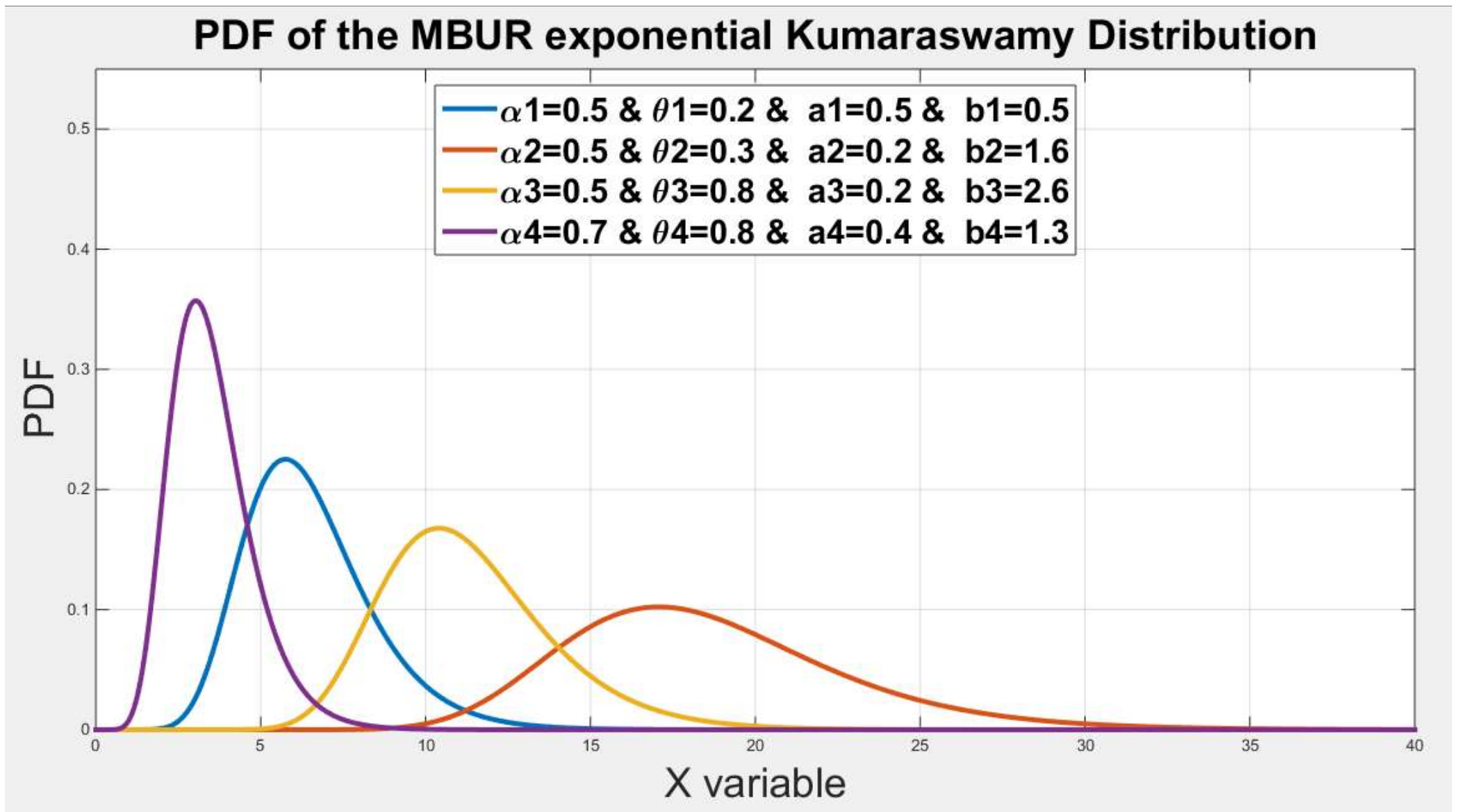

Fig. 7 Shows the PDF of the MBUR exponential {Kumaraswamy} distribution

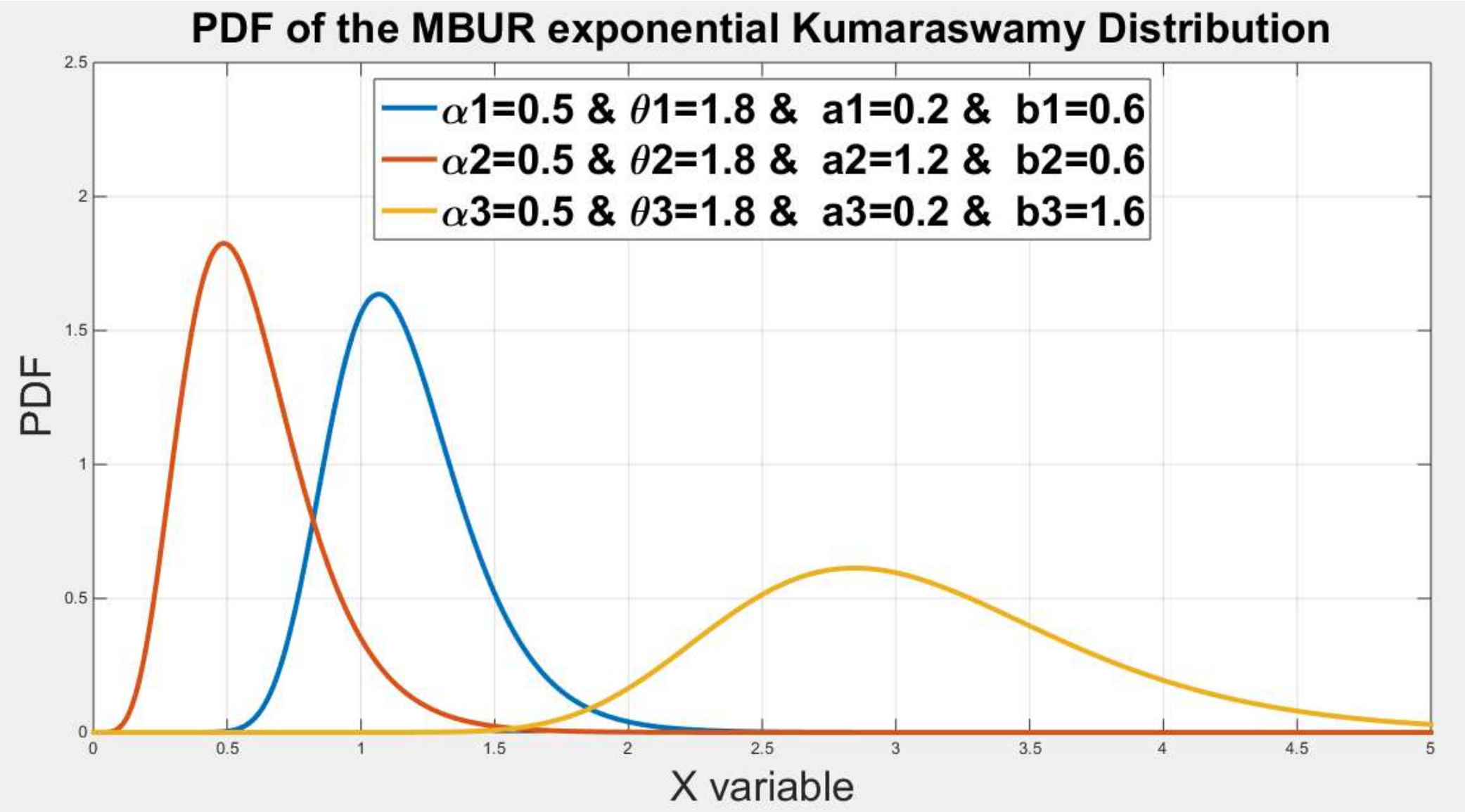

Fig. 8 Shows the PDF of the MBUR exponential {Kumaraswamy} distribution

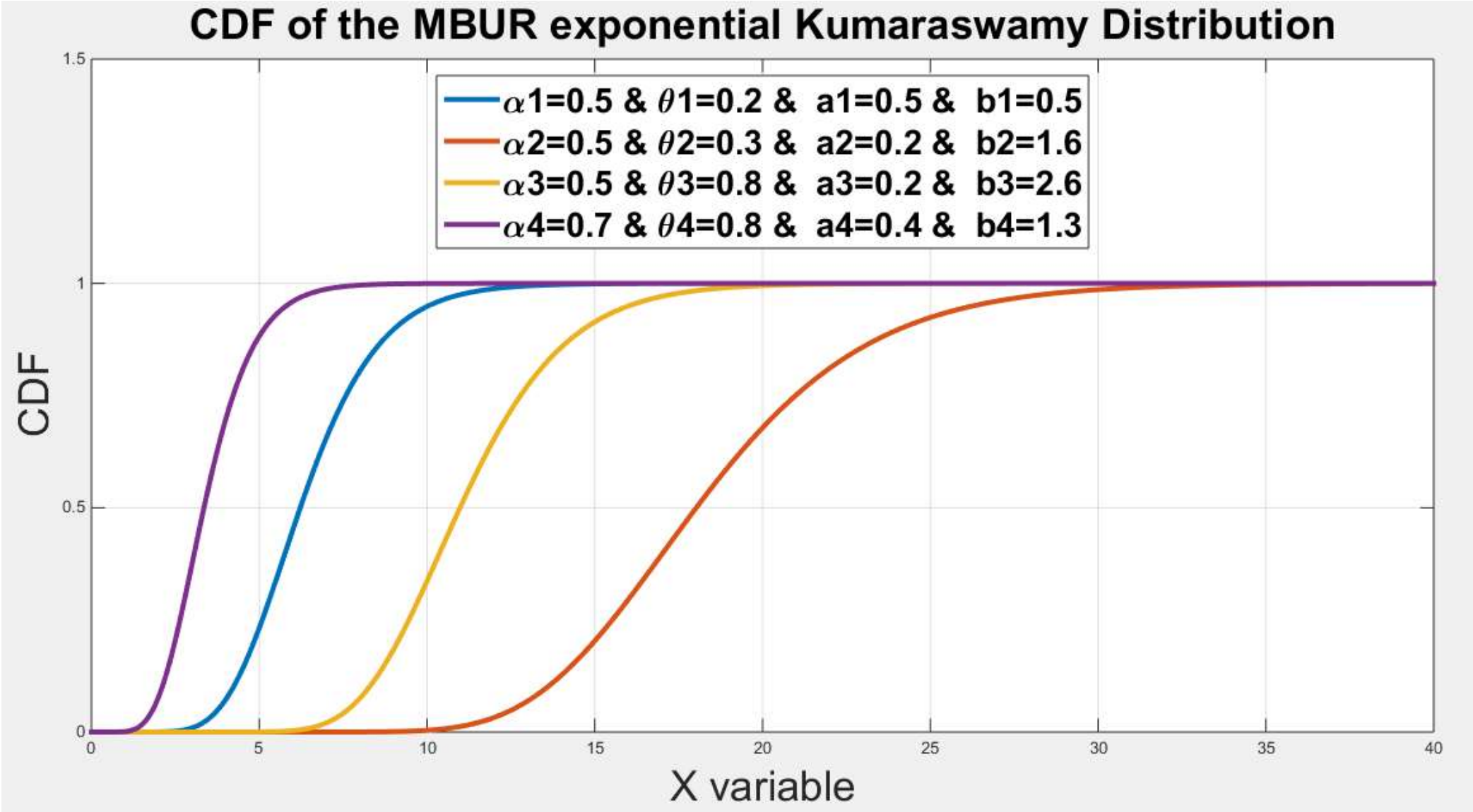

Fig. 9 shows the CDF of the MBUR exponential {Kumaraswamy} distribution

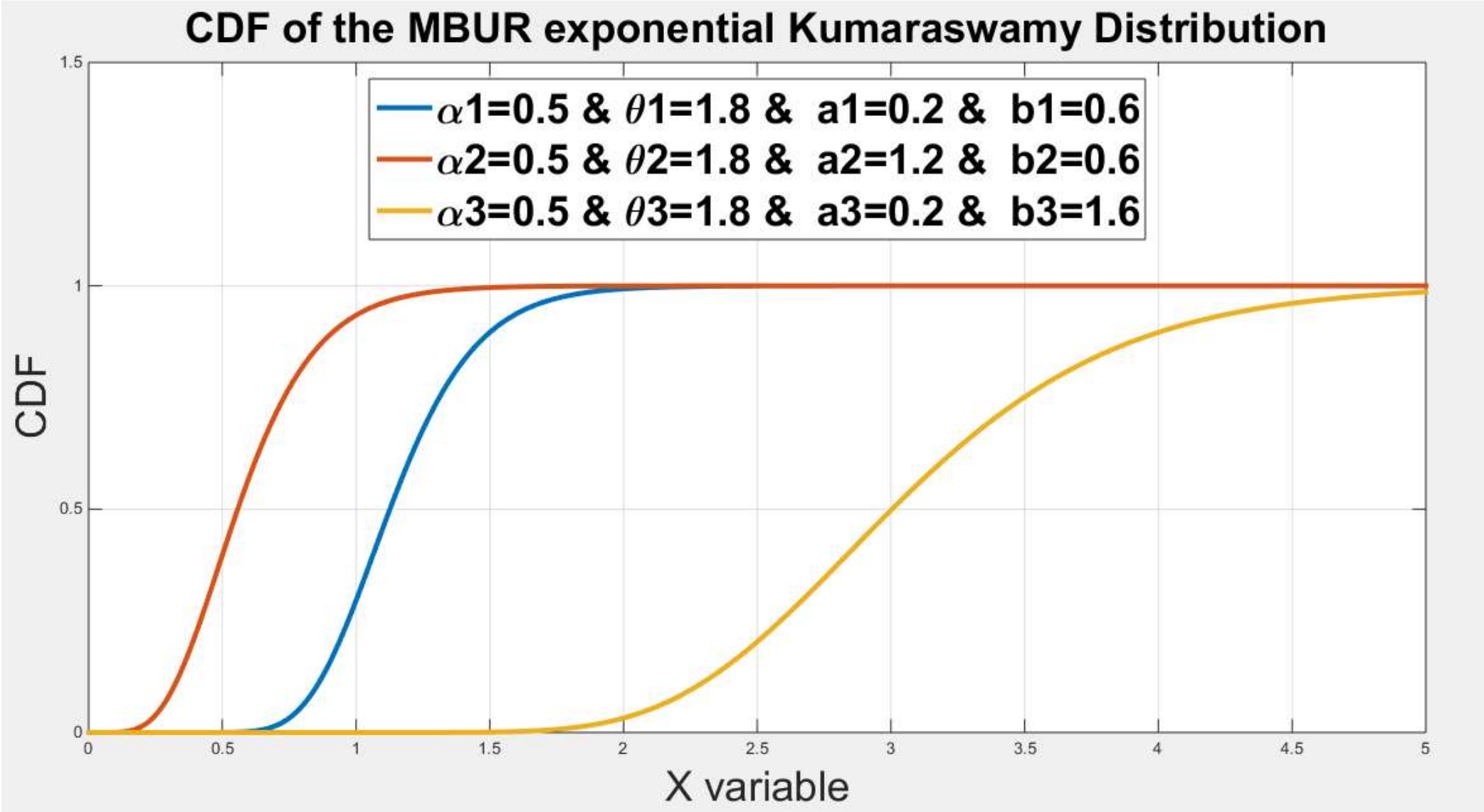

Fig. 10 shows the CDF of the MBUR exponential {Kumaraswamy} distribution

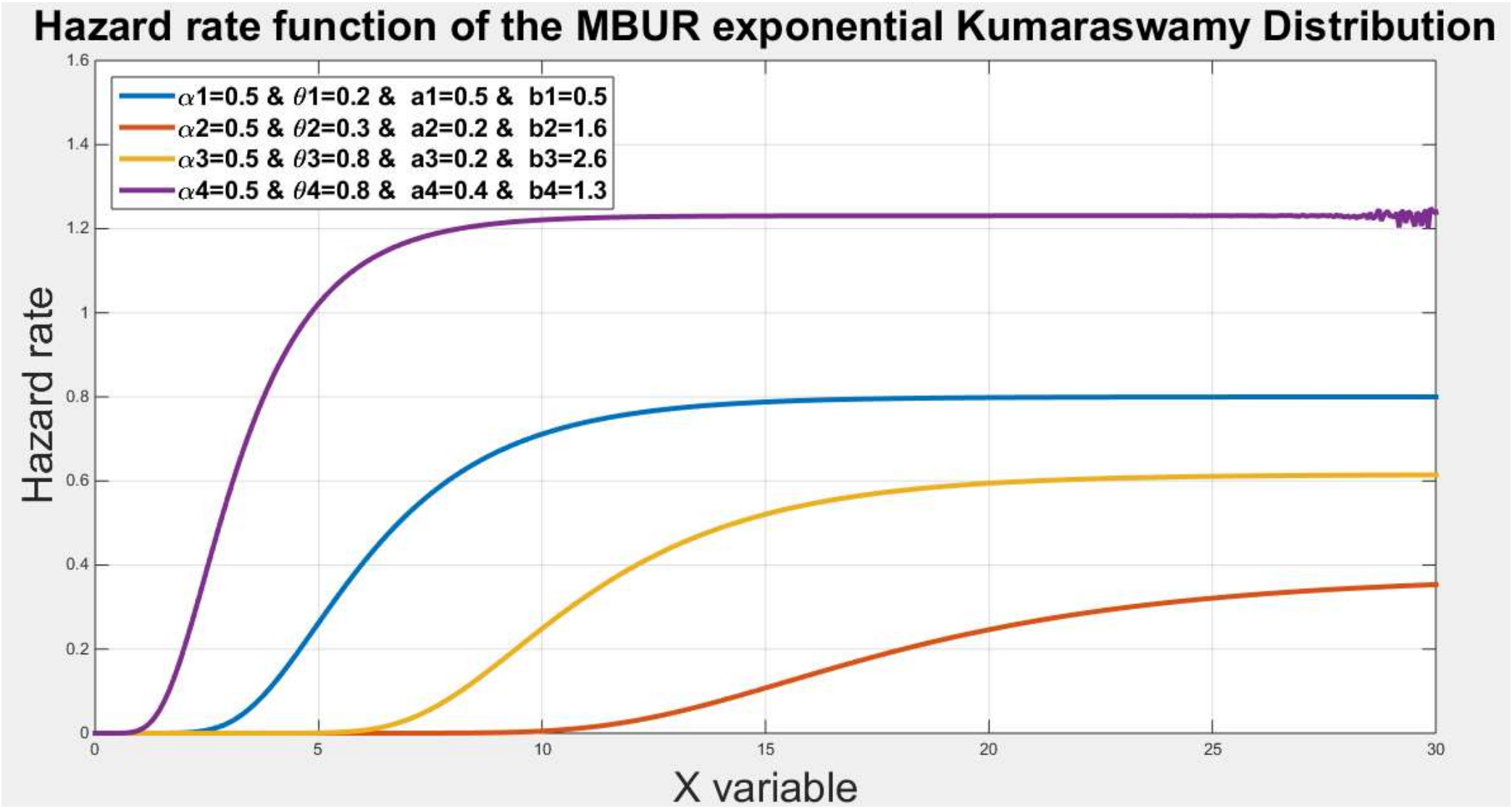

Fig. 11 shows the hazard rate function of the MBUR exponential {Kumaraswzmy } distribution

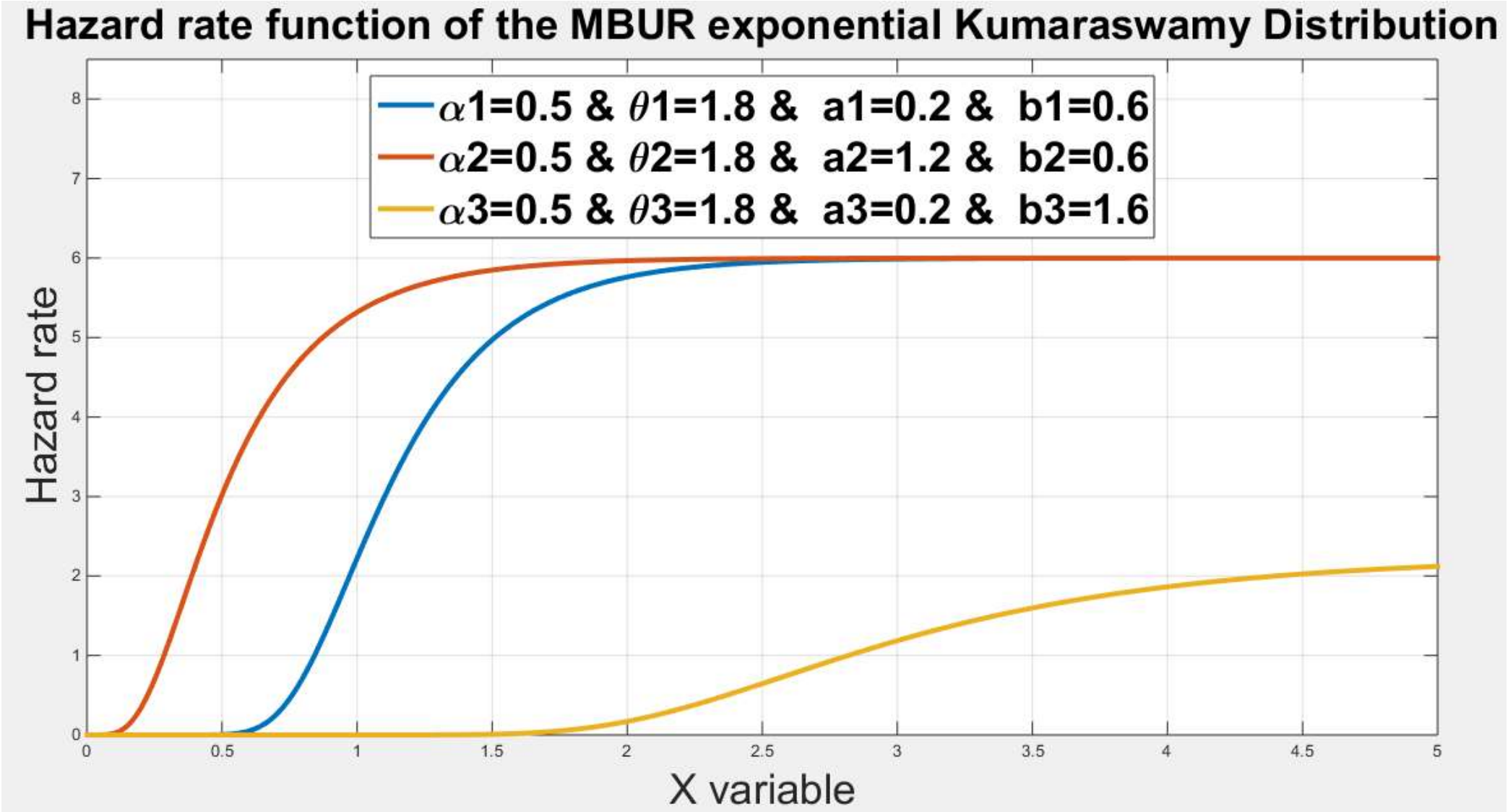

Fig. 12 shows the hazard rate function of the MBUR exponential {Kumaraswamy} distribution

# Section 3: MBUR-Exponential {Topp Leone}

## 3.1. Methodology and Derivation of the Distribution

Applying the T-X{Y} methodology as in subsection 2.1 with replacement of Topp Leone distribution as the Y distribution (link distribution)

The link Topp Leone distribution has the following PDF, CDF, and quantile functions respectively as:

$$f_Y(y) = 2\,\lambda(y)^{\lambda-1}\,(2-y)^{\lambda-1}(1-y),\ \ 0 < y < 1\ \ \&\ \lambda > 0$$

$$F_Y(y) = y^{\lambda}(2-y)^{\lambda}$$

$$Q(u;\theta) = 1 - \left(1 - u^{\frac{1}{\lambda}}\right)^{0.5}$$

utilize the following steps to acquire the new distribution:

**Step 1**: substitute the CDF function of the exponential into the quantile function of the Topp Leone

$$T = \left\{1 - \left(1 - \left(1 - e^{-\theta x}\right)^{\frac{1}{\lambda}}\right)^{0.5}\right\}$$

**Step2**: substitute this T into the pdf of the MBUR

$$f_T(t) = \frac{6}{\alpha^2}\left(1 - \left(1 - \left(1 - e^{-\theta x}\right)^{\frac{1}{\lambda}}\right)^{0.5}\right)^{\frac{2}{\alpha^2}-1}\left(1 - \left(1 - \left(1 - \left(1 - e^{-\theta x}\right)^{\frac{1}{\lambda}}\right)^{0.5}\right)^{\frac{1}{\alpha^2}}\right),$$

$$0 < t < 1 \;\; \& \alpha > 0$$

**Step 3**: substitute into the pdf of the new distribution:

$$g_{X_g}(x) = f_X(x)\,\frac{r\left(Q_Y\left(F_X(x)\right)\right)}{f_Y\left(Q_Y\left(F_X(x)\right)\right)} = f_X(x)\,\frac{f_T\left(Q_Y\left(F_X(x)\right)\right)}{f_Y\left(Q_Y\left(F_X(x)\right)\right)}$$

$$= \frac{\theta e^{-\theta x}\frac{6}{\alpha^2}\left(1 - \left(1 - \left(1 - e^{-\theta x}\right)^{\frac{1}{\lambda}}\right)^{0.5}\right)^{\frac{2}{\alpha^2}-1}\left(1 - \left(1 - \left(1 - \left(1 - e^{-\theta x}\right)^{\frac{1}{\lambda}}\right)^{0.5}\right)^{\frac{1}{\alpha^2}}\right)}{2\,\lambda\left(1 - \left(1 - \left(1 - e^{-\theta}\;\right)^{\frac{1}{\lambda}}\right)^{0.5}\right)^{\lambda-1}\left(2 - 1 + \left(1 - \left(1 - e^{-\theta}\;\right)^{\frac{1}{\lambda}}\right)^{0.5}\right)^{\lambda-1}\left(1 - 1 + \left(1 - \left(1 - e^{-\theta x}\right)^{\frac{1}{\lambda}}\right)^{0.5}\right)}$$

$$= \frac{\theta e^{-\theta x}\frac{6}{\alpha^2}\left(1 - \left(1 - \left(1 - e^{-\theta x}\right)^{\frac{1}{\lambda}}\right)^{0.5}\right)^{\frac{2}{\alpha^2}-1}\left(1 - \left(1 - \left(1 - \left(1 - e^{-\theta x}\right)^{\frac{1}{\lambda}}\right)^{0.5}\right)^{\frac{1}{\alpha^2}}\right)}{2\,\lambda\left(1 - \left(1 - \left(1 - e^{-\theta x}\right)^{\frac{1}{\lambda}}\right)^{0.5}\right)^{\lambda-1}\left(2 - 1 + \left(1 - \left(1 - e^{-\theta x}\right)^{\frac{1}{\lambda}}\right)^{0.5}\right)^{\lambda-1}\left(1 - 1 + \left(1 - \left(1 - e^{-\theta x}\right)^{\frac{1}{\lambda}}\right)^{0.5}\right)}$$

$$= \frac{\theta e^{-\theta x}\frac{6}{\alpha^2}\left(1 - \left(1 - \left(1 - e^{-\theta x}\right)^{\frac{1}{\lambda}}\right)^{0.5}\right)^{\frac{2}{\alpha^2}-1}\left(1 - \left(1 - \left(1 - \left(1 - e^{-\theta x}\right)^{\frac{1}{\lambda}}\right)^{0.5}\right)^{\frac{1}{\alpha^2}}\right)}{2\,\lambda\left(1 - \left(1 - \left(1 - e^{-\theta x}\right)^{\frac{1}{\lambda}}\right)^{0.5}\right)^{\lambda-1}\left(1 + \left(1 - \left(1 - e^{-\theta x}\right)^{\frac{1}{\lambda}}\right)^{0.5}\right)^{\lambda-1}\left(\left(1 - \left(1 - e^{-\theta x}\right)^{\frac{1}{\lambda}}\right)^{0.5}\right)}$$

$$= \frac{3\theta}{\lambda\alpha^2}e^{-\theta x}\,\frac{\left(1 - \left(1 - \left(1 - e^{-\theta x}\right)^{\frac{1}{\lambda}}\right)^{0.5}\right)^{\frac{2}{\alpha^2}-\lambda}\left(1 - \left(1 - \left(1 - \left(1 - e^{-\theta x}\right)^{\frac{1}{\lambda}}\right)^{0.5}\right)^{\frac{1}{\alpha^2}}\right)}{\left(1 + \left(1 - \left(1 - e^{-\theta x}\right)^{\frac{1}{\lambda}}\right)^{0.5}\right)^{\lambda-1}\left(\left(1 - \left(1 - e^{-\theta x}\right)^{\frac{1}{\lambda}}\right)^{0.5}\right)}$$

$$g_{X_g}(x) = \frac{3\theta}{\lambda\alpha^2} e^{-\theta} \left(1-\left(1-\left(1-e^{-\theta x}\right)^{\frac{1}{\lambda}}\right)^{0.5}\right)^{\frac{2}{\alpha^2}-\lambda} \left(1-\left(1-\left(1-\left(1-e^{-\theta x}\right)^{\frac{1}{\lambda}}\right)^{0.5}\right)^{\frac{1}{\alpha^2}}\right)$$

$$\left(1+\left(1-\left(1-e^{-\theta x}\right)^{\frac{1}{\lambda}}\right)^{0.5}\right)^{1-\lambda} \left(\left(1-\left(1-e^{-\theta x}\right)^{\frac{1}{\lambda}}\right)^{-0.5}\right)$$

The CDF of the new distribution is attained by substituting the T variable into the CDF of the MBUR $F_T(t) = 3(t)^{\frac{2}{\alpha^2}} - 2(t)^{\frac{3}{\alpha^2}}$, thus

$$G_{X_g}(x) = F_T(t) = 3\left( 1-\left(1-\left(1-e^{-\theta}\ \right)^{\frac{1}{\lambda}}\right)^{0.5}\right)^{\frac{2}{\alpha^2}} - 2\left(1-\left(1-\left(1-e^{-\theta x}\right)^{\frac{1}{\lambda}}\right)^{0.5}\ \right)^{\frac{3}{\alpha^2}}$$

## 3.2 Properties of The Distribution

### 3.2.1 The hazard function

**Proposition**: hazard function is defined as $hr = \frac{g_{X_g}(x)}{1-G_{X_g}(x)}$

$$= \frac{\frac{3\theta}{\lambda\alpha^2} e^{-\theta x} \left(1-\left(1-\left(1-e^{-\theta x}\right)^{\frac{1}{\lambda}}\right)^{0.5}\right)^{\frac{2}{\alpha^2}-\lambda} \left(1-\left(1-\left(1-\left(1-e^{-\theta x}\right)^{\frac{1}{\lambda}}\right)^{0.5}\right)^{\frac{1}{\alpha^2}}\right)\left(1+\left(1-\left(1-e^{-\theta x}\right)^{\frac{1}{\lambda}}\right)^{0.5}\right)^{1-\lambda} \left(\left(1-\left(1-e^{-\theta x}\right)^{\frac{1}{\lambda}}\right)^{-0.5}\right)}{1-3\left( 1-\left(1-\left(1-e^{-\theta x}\right)^{\frac{1}{\lambda}}\right)^{0.5}\right)^{\frac{2}{\alpha^2}} + 2\left(1-\left(1-\left(1-e^{-\theta x}\right)^{\frac{1}{\lambda}}\right)^{0.5}\ \right)^{\frac{3}{\alpha^2}}}$$

### 3.2.2 The quantile function:

**Theorem**: the quantile function is defined as:

$$Q = \frac{-1}{\theta}\ln\left(1-\left\{\left(2-\left\{-.5\left(cos\left[\frac{cos^{-1}(1-2u)}{3}\right]-\sqrt{3}\,sin\left[\frac{cos^{-1}(1-2u)}{3}\right]\right)+.5\right\}^{\alpha^2}\right)^{\lambda}\left(\left\{-.5\left(cos\left[\frac{cos^{-1}(1-2u)}{3}\right]-\sqrt{3}\,sin\left[\frac{cos^{-1}(1-2u)}{3}\right]\right)+.5\right\}^{\alpha^2}\ \right)^{\lambda}\right\}\right)$$

**proof:** By definition, CDF of the new distribution is defined as

$G_{X_g}(x) = R\left(Q_Y\left(F_X(x)\right)\right) = F_T\ \left(Q_Y\left(F_X(x)\right)\right)$ applying the following steps

**Step 1:** Take the inverse of both sides, use the inverse function of the MBUR

$$u = F_T\ \left(Q_Y\left(F_X(x)\right)\right)$$

$$F_T^{-1}(u) = F_T^{-1}\, F_T\ \left(Q_Y\left(F_X(x)\right)\right)$$

$$t = \left\{-.5\left(cos\left[\frac{cos^{-1}(1-2u)}{3}\right]-\sqrt{3}\,sin\left[\frac{cos^{-1}(1-2u)}{3}\right]\right)+.5\right\}^{\alpha^2} = Q_Y\left(F_X(x)\right)$$

**Step 2:** Take the CDF of both side: substitute this (t) into the CDF of the quantile distribution

$$F_T^{-1}(u) = \left(Q_Y\left(F_X(x)\right)\right)$$

$$F_Y[F_T^{-1}(u)] = F_Y\left(Q_Y\left(F_X(x)\right)\right)$$

$$F_Y[F_T^{-1}(u)] = \left(\left(F_X(x)\right)\right)$$

Where $F_Y(y) = y^{\lambda}(2-y)^{\lambda}$ so

$$F_Y(t) = t^{\lambda}(2-t)^{\lambda}$$

$$F_Y(y) = \left(\left\{-.5\left(\cos\left[\frac{\cos^{-1}(1-2u)}{3}\right] - \sqrt{3}\sin\left[\frac{\cos^{-1}(1-2u)}{3}\right]\right) + .5\right\}^{\alpha^2}\right)^{\lambda}$$

$$\left(2 - \left(\left\{-.5\left(\cos\left[\frac{\cos^{-1}(1-2u)}{3}\right] - \sqrt{3}\sin\left[\frac{\cos^{-1}(1-2u)}{3}\right]\right) + .5\right\}^{\alpha^2}\right)\right)^{\lambda}$$

**Step3:** Take the inverse once again on both side, i.e. substitute the CDF from the previous step into the quantile of the exponential

$$F_X^{-1}\{F_Y[F_T^{-1}(u)]\} = F_X^{-1}\left(\left(F_X(x)\right)\right)$$

$$Q(u) = x = F_X^{-1}\{F_Y[F_T^{-1}(u)]\}$$

$$Q(u;\theta,\alpha,\lambda) = \frac{-1}{\theta}\ln(1-u)$$

$$Q = \frac{-1}{\theta}\ln\left(1 - F_Y(t)\right)$$

$$Q = \frac{-1}{\theta}\ln\left(1 - \left\{\left(2 - \left\{-.5\left(\cos\left[\frac{\cos^{-1}(1-2u)}{3}\right] - \sqrt{3}\sin\left[\frac{\cos^{-1}(1-2u)}{3}\right]\right) + .5\right\}^{\alpha^2}\right)^{\lambda}\left(\left\{-.5\left(\cos\left[\frac{\cos^{-1}(1-2u)}{3}\right] - \sqrt{3}\sin\left[\frac{\cos^{-1}(1-2u)}{3}\right]\right) + .5\right\}^{\alpha^2}\right)^{\lambda}\right\}\right)$$

### 3.2.3 The $r^{th}$ raw moment

**Theorem**: the raw moment can be defined as

$$E(x^r) = \frac{3\Gamma(r+1)}{\lambda\alpha^2\theta^r}\left(\frac{w_{q,p,r,s,t,f,g,h}}{(r+t+h+1)^{(r+1)}} - \frac{w_{q,p,r,s,t,f,g,h,k,m,n}}{(r+t+h+n+1)^{(r+1)}}\right)$$

**Proof:**

$g_{X_g}(x)$ can be rewritten using the generalized binomial expansion for the following terms:

$$\frac{3\theta}{\lambda\alpha^2}e^{-\theta x}\left(1-\left(1-\left(1-e^{-\theta x}\right)^{\frac{1}{\lambda}}\right)^{0.5}\right)^{\frac{2}{\alpha^2}-\lambda}\left(1-\left(1-\left(1-\left(1-e^{-\theta x}\right)^{\frac{1}{\lambda}}\right)^{0.5}\right)^{\frac{1}{\alpha^2}}\right)\left(1+\left(1-\left(1-e^{-\theta x}\right)^{\frac{1}{\lambda}}\right)^{0.5}\right)^{1-\lambda}\left(\left(1-\left(1-e^{-\theta x}\right)^{\frac{1}{\lambda}}\right)^{-0.5}\right)$$

First term:

$$\left(1-\left(1-\left(1-e^{-\theta x}\right)^{\frac{1}{\lambda}}\right)^{0.5}\right)^{\frac{2}{\alpha^2}-\lambda} = \sum_{q=0}^{\infty}(-1)^q \binom{\frac{2}{\alpha^2}-\lambda}{q}\left(1-\left(1-e^{-\theta x}\right)^{\frac{1}{\lambda}}\right)^{\frac{q}{2}}$$

$$=\sum_{q=0}^{\infty}(-1)^q \binom{\frac{2}{\alpha^2}-\lambda}{q}\sum_{p=0}^{\infty}(-1)^p \binom{\frac{q}{2}}{p}\left(1-e^{-\theta x}\right)^{\frac{p}{\lambda}} = \sum_{q=0}^{\infty}(-1)^q \binom{\frac{2}{\alpha^2}-\lambda}{q}\sum_{p=0}^{\infty}(-1)^p \binom{\frac{q}{2}}{p}\sum_{r=0}^{\infty}(-1)^r \binom{\frac{p}{\lambda}}{r} e^{-\theta r x}$$

$$=\sum_{q=0}^{\infty}\sum_{p=0}^{\infty}\sum_{r=0}^{\infty}(-1)^{q+p+r}\binom{\frac{2}{\alpha^2}-\lambda}{q}\binom{\frac{q}{2}}{p}\binom{\frac{p}{\lambda}}{r}e^{-\theta r x}$$

Second term:

$$\left(1-\left(1-\left(1-\left(1-e^{-\theta}\ \right)^{\frac{1}{\lambda}}\right)^{0.5}\right)^{\frac{1}{\alpha^2}}\right) = 1-\sum_{k=0}^{\infty}(-1)^k \binom{\frac{1}{\alpha^2}}{k}\left(1-\left(1-e^{-\theta}\ \right)^{\frac{1}{\lambda}}\right)^{\frac{k}{2}}$$

$$=1-\sum_{k=0}^{\infty}(-1)^k \binom{\frac{1}{\alpha^2}}{k}\sum_{m=0}^{\infty}(-1)^m \binom{\frac{k}{2}}{m}\left(1-e^{-\theta x}\right)^{\frac{m}{\lambda}}$$

$$=1-\sum_{k=0}^{\infty}(-1)^k \binom{\frac{1}{\alpha^2}}{k}\sum_{m=0}^{\infty}(-1)^m \binom{\frac{k}{2}}{m}\sum_{n=0}^{\infty}(-1)^n \binom{\frac{m}{\lambda}}{n} e^{-\theta n x}$$

$$=\left(1-\sum_{k=0}^{\infty}\sum_{m=0}^{\infty}\sum_{n=0}^{\infty}(-1)^{k+m+n}\binom{\frac{1}{\alpha^2}}{k}\binom{\frac{k}{2}}{m}\binom{\frac{m}{\lambda}}{n} e^{-\theta n x}\right)$$

Third term:

$$\left(\left(1-\left(1-e^{-\theta x}\right)^{\frac{1}{\lambda}}\right)^{-0.5}\right) = \sum_{s=0}^{\infty}(-1)^s \binom{-0.5}{s}\left(1-e^{-\theta x}\right)^{\frac{s}{\lambda}}$$

$$=\sum_{s=0}^{\infty}(-1)^s \binom{-0.5}{s}\sum_{t=0}^{\infty}(-1)^t \binom{\frac{s}{\lambda}}{t}e^{-\theta t x} = \sum_{s=0}^{\infty}\sum_{t=0}^{\infty}(-1)^{s+t}\binom{-0.5}{s}\binom{\frac{s}{\lambda}}{t}e^{-\theta t x}$$

Fourth term:

$$\left(1+\left(1-\left(1-e^{-\theta x}\right)^{\frac{1}{\lambda}}\right)^{0.5}\right)^{1-\lambda} = \sum_{f=0}^{\infty}\binom{1-\lambda}{f}\left(1-\left(1-e^{-\theta x}\right)^{\frac{1}{\lambda}}\right)^{\frac{f}{2}}$$

$$=\sum_{f=0}^{\infty}\binom{1-\lambda}{f}\sum_{g=0}^{\infty}(-1)^g \binom{\frac{f}{2}}{g}\left(1-e^{-\theta x}\right)^{\frac{g}{\lambda}}$$

$$=\sum_{f=0}^{\infty}\binom{1-\lambda}{f}\sum_{g=0}^{\infty}(-1)^g \binom{\frac{f}{2}}{g}\sum_{h=0}^{\infty}(-1)^h \binom{\frac{g}{\lambda}}{h}e^{-\theta h x}$$

$$= \sum_{f=0}^{\infty}\sum_{g=0}^{\infty}\sum_{h=0}^{\infty}(-1)^{g+h}\binom{1-\lambda}{f}\binom{\frac{f}{2}}{g}\binom{\frac{g}{\lambda}}{h}e^{-\theta hx}$$

Assemble the above expansions to define $g_{X_g}(x) =$

$$\sum_{q=0}^{\infty}\sum_{p=0}^{\infty}\sum_{r=0}^{\infty}(-1)^{q+p+r}\binom{\frac{2}{\alpha^2}-\lambda}{q}\binom{\frac{q}{2}}{p}\binom{\frac{p}{\lambda}}{r}e^{-\theta}\sum_{s=0}^{\infty}\sum_{t=0}^{\infty}(-1)^{s+t}\binom{-0.5}{s}\binom{\frac{s}{\lambda}}{t}e^{-\theta tx}\sum_{f=0}^{\infty}\sum_{g=0}^{\infty}\sum_{h=0}^{\infty}(-1)^{g+h}\binom{1-\lambda}{f}\binom{\frac{f}{2}}{g}\binom{\frac{g}{\lambda}}{h}e^{-\theta hx}$$

$$\left(1-\sum_{k=0}^{\infty}\sum_{m=0}^{\infty}\sum_{n=0}^{\infty}(-1)^{k+m+n}\binom{\frac{1}{\alpha^2}}{k}\binom{\frac{k}{2}}{m}\binom{\frac{m}{\lambda}}{n}e^{-\theta n}\right)$$

$$g_{X_g}(x) = \frac{3\theta}{\lambda\alpha^2}e^{-\theta x}$$

$$\sum_{q=0}^{\infty}\sum_{p=0}^{\infty}\sum_{r=0}^{\infty}\sum_{s=0}^{\infty}\sum_{t=0}^{\infty}\sum_{f=0}^{\infty}\sum_{g=0}^{\infty}\sum_{h=0}^{\infty}(-1)^{q+p+r+s+t+g+h}\binom{\frac{2}{\alpha^2}-\lambda}{q}\binom{\frac{q}{2}}{p}\binom{\frac{p}{\lambda}}{r}\binom{-0.5}{s}\binom{\frac{s}{\lambda}}{t}\binom{1-\lambda}{f}\binom{\frac{f}{2}}{g}\binom{\frac{g}{\lambda}}{h}e^{-\theta x(r+t+h)}$$

$$\left(1-\sum_{k=0}^{\infty}\sum_{m=0}^{\infty}\sum_{n=0}^{\infty}(-1)^{k+m+n}\binom{\frac{1}{\alpha^2}}{k}\binom{\frac{k}{2}}{m}\binom{\frac{m}{\lambda}}{n}e^{-\theta nx}\right)$$

Finally, $g_{X_g}(x) = \frac{3\theta}{\lambda\alpha^2}e^{-\theta x}w_{q,p,r,s,t,f,g,h}\,e^{-\theta x(r+t+h)}\left(1-w_{k,m,n}\,e^{-\theta nx}\right)$

Where

$$w_{q,p,r,s,t,f,g,h} = \sum_{q=0}^{\infty}\sum_{p=0}^{\infty}\sum_{r=0}^{\infty}\sum_{s=0}^{\infty}\sum_{t=0}^{\infty}\sum_{f=0}^{\infty}\sum_{g=0}^{\infty}\sum_{h=0}^{\infty}(-1)^{q+p+r+s+t+g+h}\binom{\frac{2}{\alpha^2}-\lambda}{q}\binom{\frac{q}{2}}{p}\binom{\frac{p}{\lambda}}{r}\binom{-0.5}{s}\binom{\frac{s}{\lambda}}{t}\binom{1-\lambda}{f}\binom{\frac{f}{2}}{g}\binom{\frac{g}{\lambda}}{h}$$

$$w_{k,m,n} = \sum_{k=0}^{\infty}\sum_{m=0}^{\infty}\sum_{n=0}^{\infty}(-1)^{k+m+n}\binom{\frac{1}{\alpha^2}}{k}\binom{\frac{k}{2}}{m}\binom{\frac{m}{\lambda}}{n}$$

By definition : $E(x^r) = \int_0^{\infty} x^r\, g_{X_g}(x)\,dx$

$$E(x^r) = \frac{3\theta}{\lambda\alpha^2}\int_0^{\infty} x^r\left\{w_{q,p,r,s,t,f,g,h}\,e^{-\theta x(r+t+h+1)}\left(1-w_{k,m,n}\,e^{-\theta nx}\right)\right\}dx$$

$$E(x^r) = \frac{3\theta}{\lambda\alpha^2}\int_0^{\infty} x^r\,w_{q,p,r,s,t,f,g,h}\,e^{-\theta x(r+t+h+1)}\,dx$$

$$-\frac{3\theta}{\lambda\alpha^2}\int_0^{\infty} x^r\,w_{q,p,r,s,t,f,g,h,k,m,n}\,e^{-\theta x(r+t+h+n+1)}\,dx$$

$$E(x^r) = \frac{3\theta}{\lambda\alpha^2}\left\{\frac{(w_{q,p,r,s,t,f,g,h})\Gamma(r+1)}{[\theta(r+t+h+1)]^{(r+1)}}-\frac{(w_{q,p,r,s,t,f,g,h,k,m,n})\Gamma(r+1)}{[\theta(r+t+h+n+1)]^{(r+1)}}\right\}$$

$$E(x^r) = \frac{3\theta}{\lambda\alpha^2}\left(\frac{\Gamma(r+1)}{\theta^{(r+1)}}\right)\left(\frac{w_{q,p,r,s,t,f,g,h}}{(r+t+h+1)^{(r+1)}}-\frac{w_{q,p,r,s,t,f,g,h,k,m,n}}{(r+t+h+n+1)^{(r+1)}}\right)$$

$$E(x^r) = \frac{3}{\lambda\alpha^2}\left(\frac{\Gamma(r+1)}{\theta^r}\right)\left(\frac{w_{q,p,r,s,t,f,g,h}}{(r+t+h+1)^{(r+1)}} - \frac{w_{q,p,r,s,t,f,g,h,k,m,n}}{(r+t+h+n+1)^{(r+1)}}\right)$$

Figures (13-18) show the PDF, CDF, hazard rate functions. The PDF can attain decreasing shapes or skewed shaped. The hazard functions can be initially increasing or decreasing then they attain a constant rate.

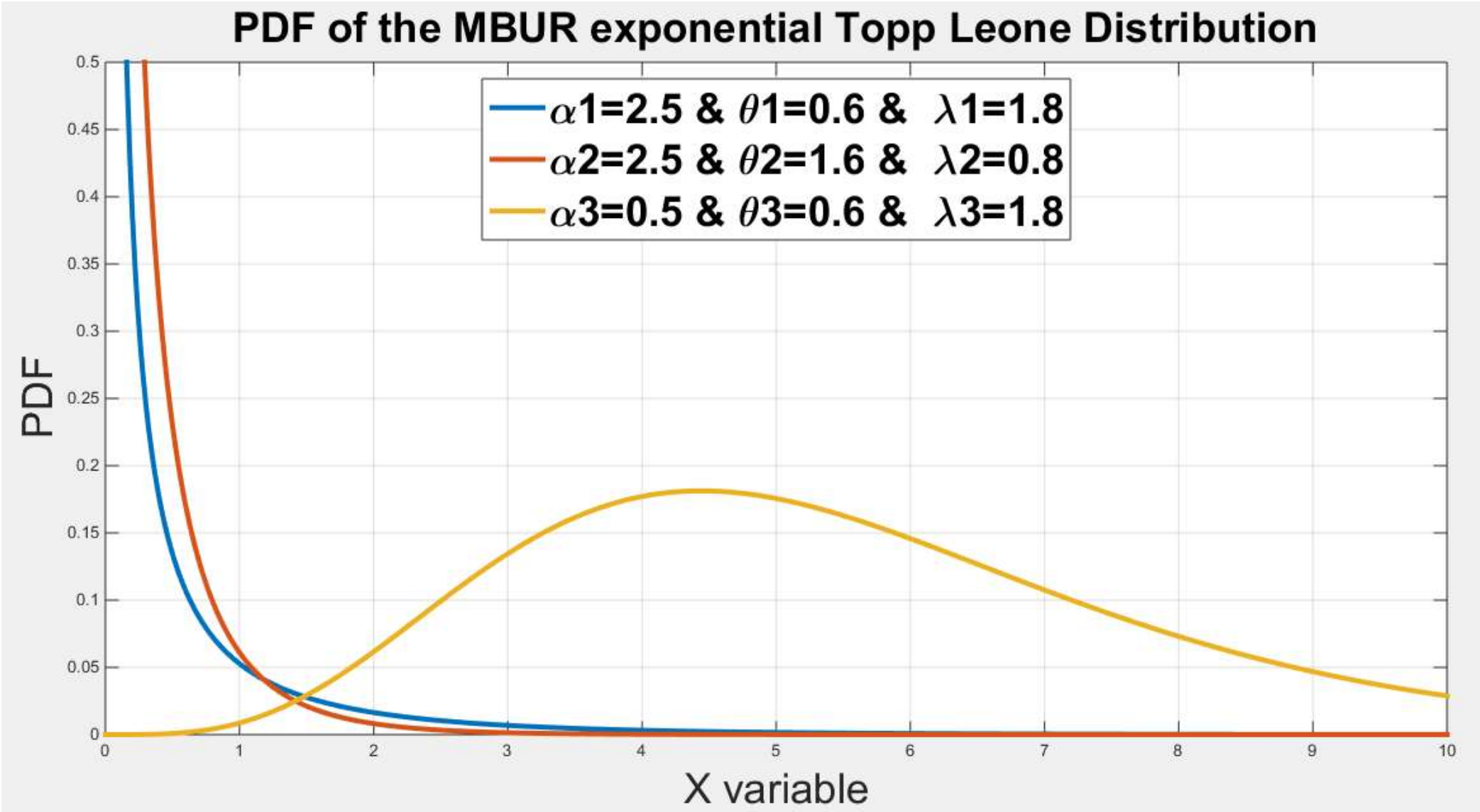

Fig. 13 shows the PDF of the MBUR exponential {Topp Leone} distribution.

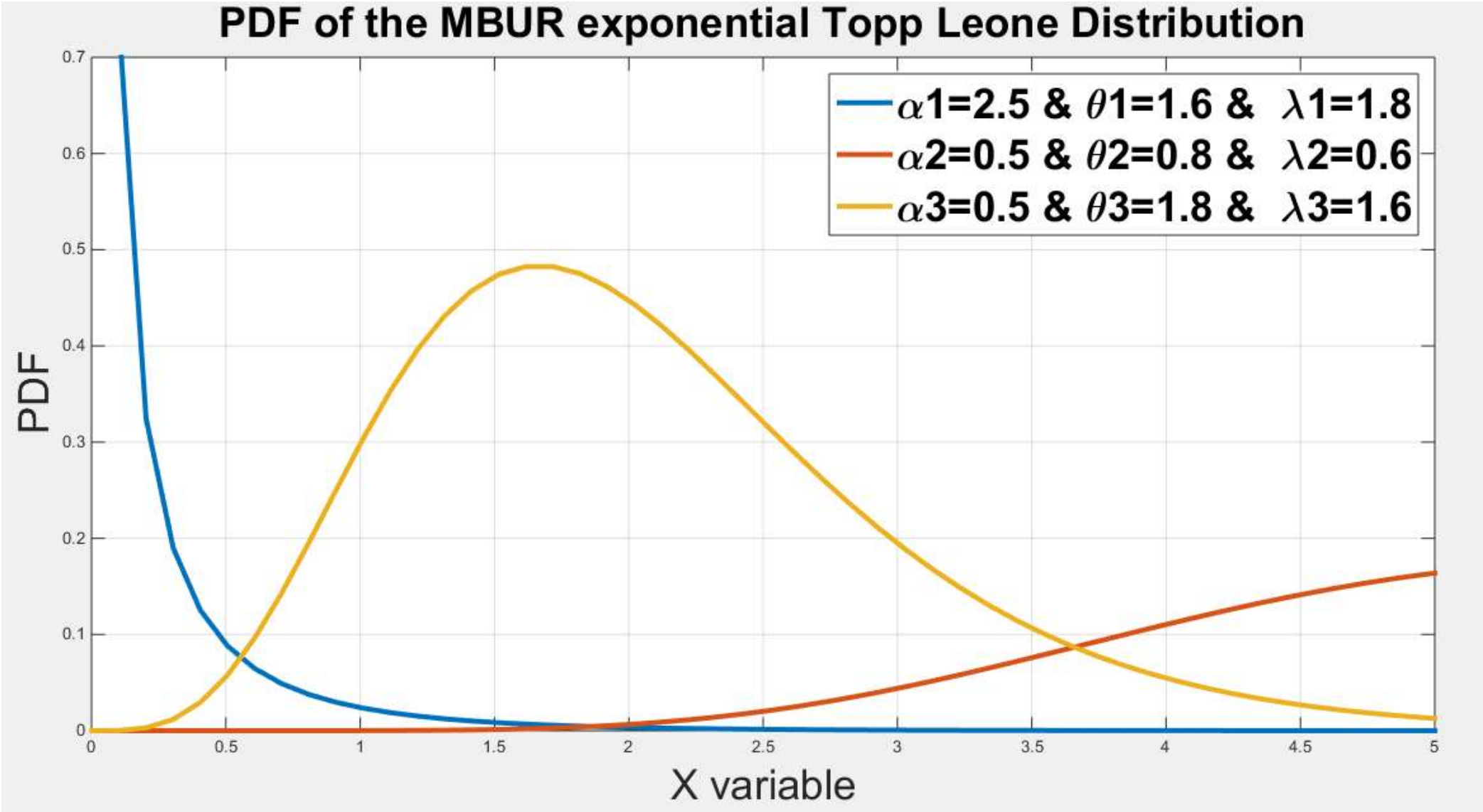

Fig. 14 shows the PDF of the MBUR exponential {Topp Leone} distribution.

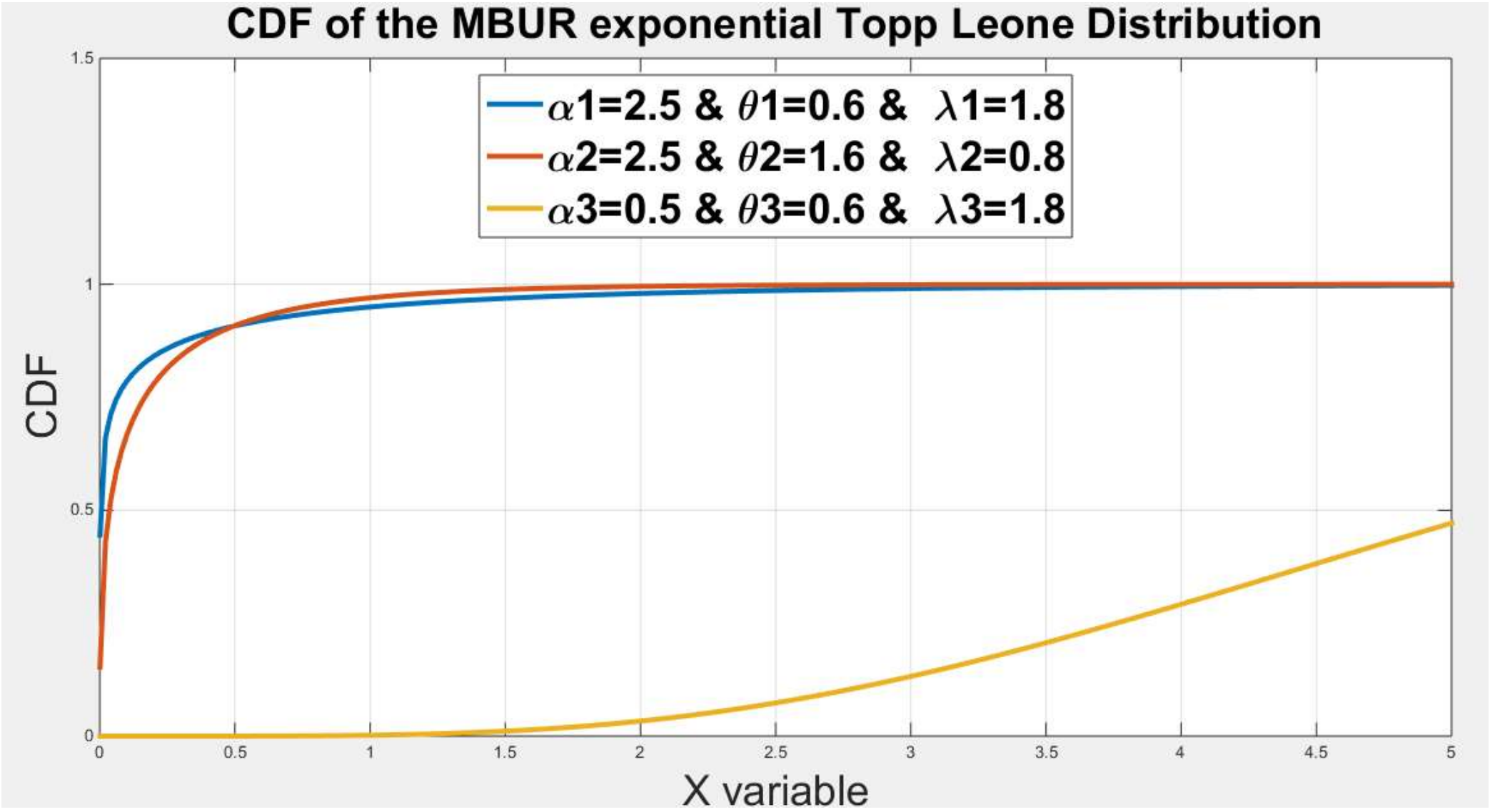

Fig. 15 shows the CDF of the MBUR exponential {Topp Leone} distribution

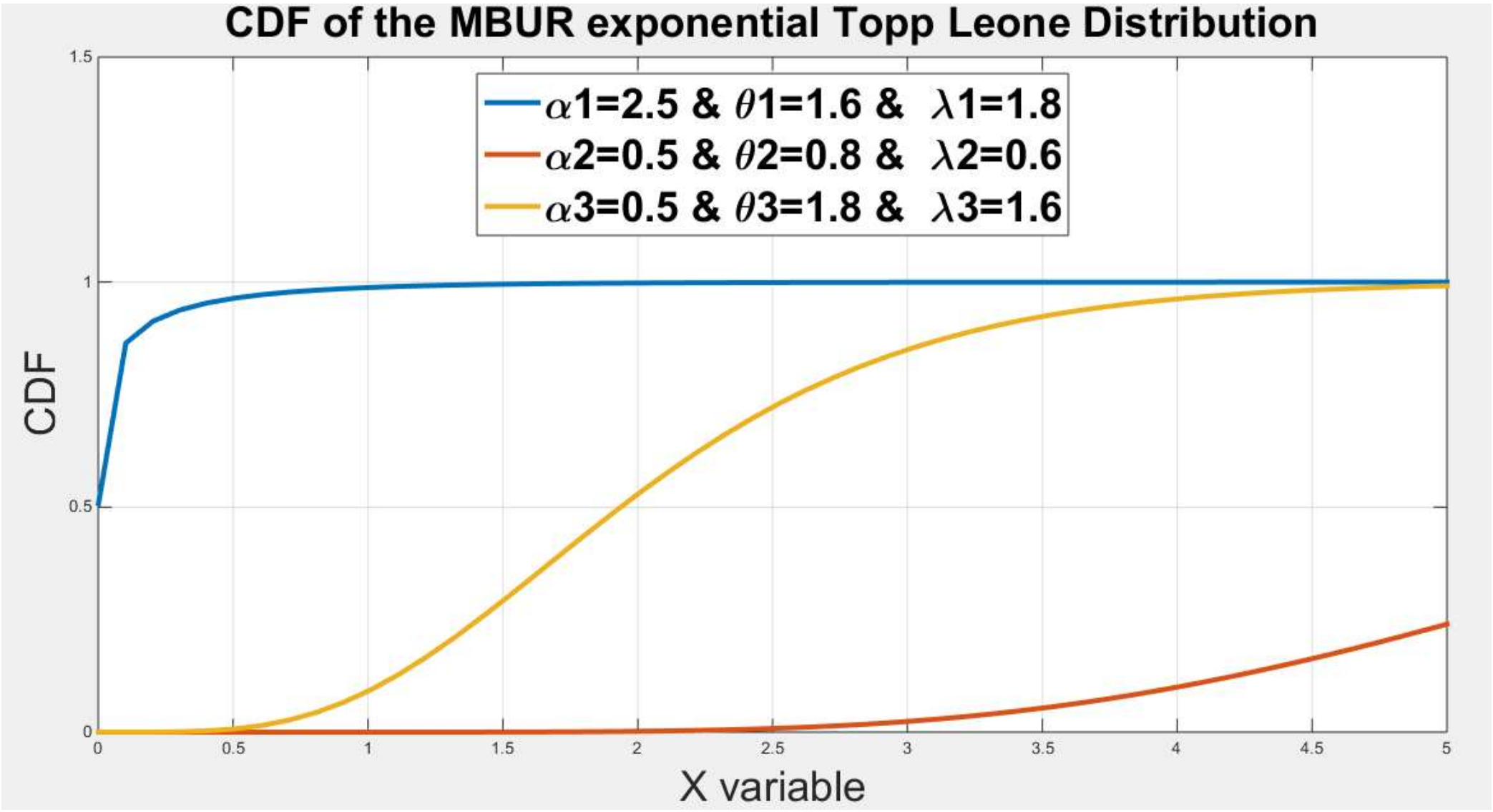

Fig. 16 shows the CDF of the MBUR exponential {Topp Leone} distribution

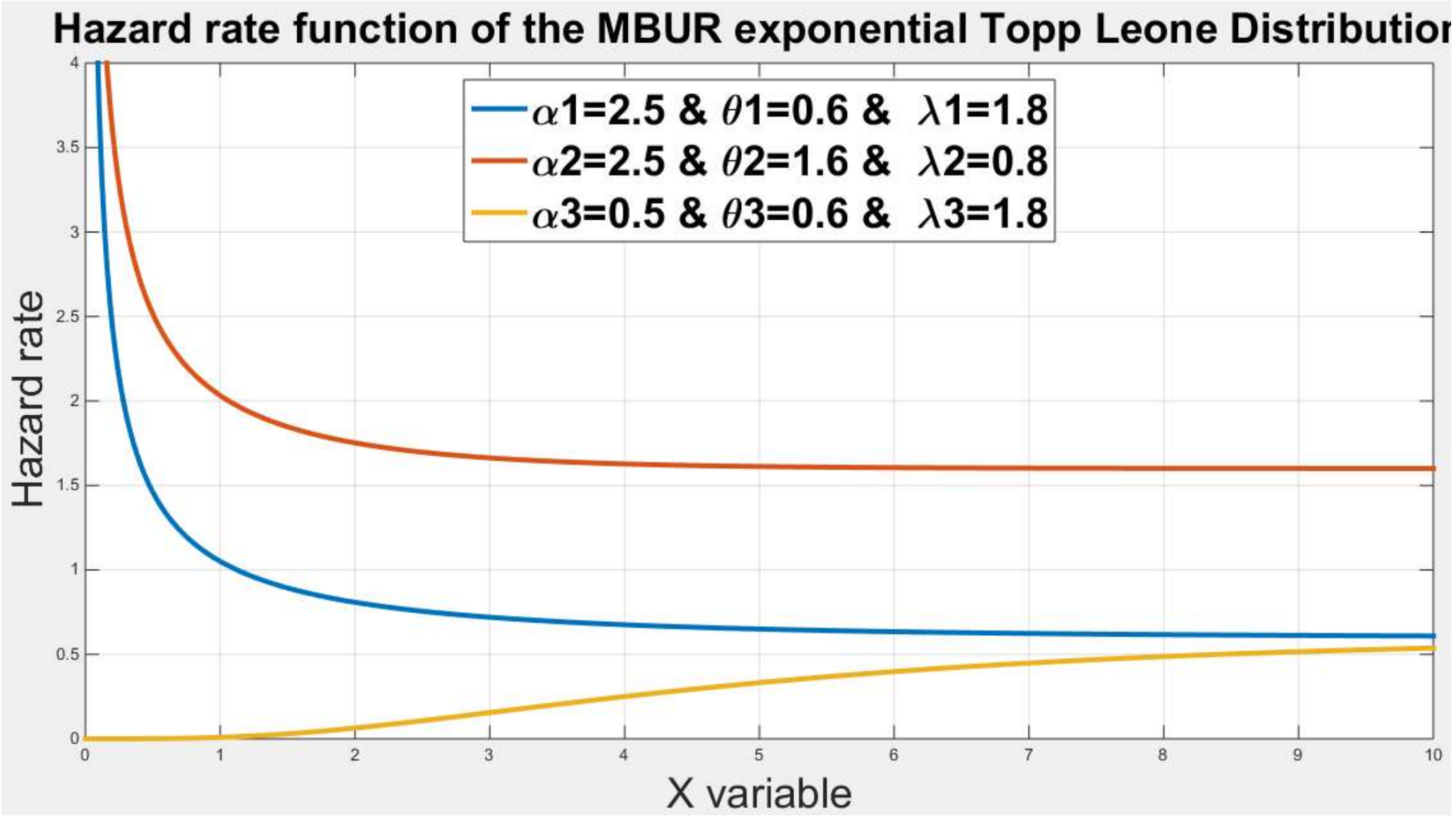

Fig. 17 shows the hazard rate function of the MBUR exponential {Topp Leone} distribution

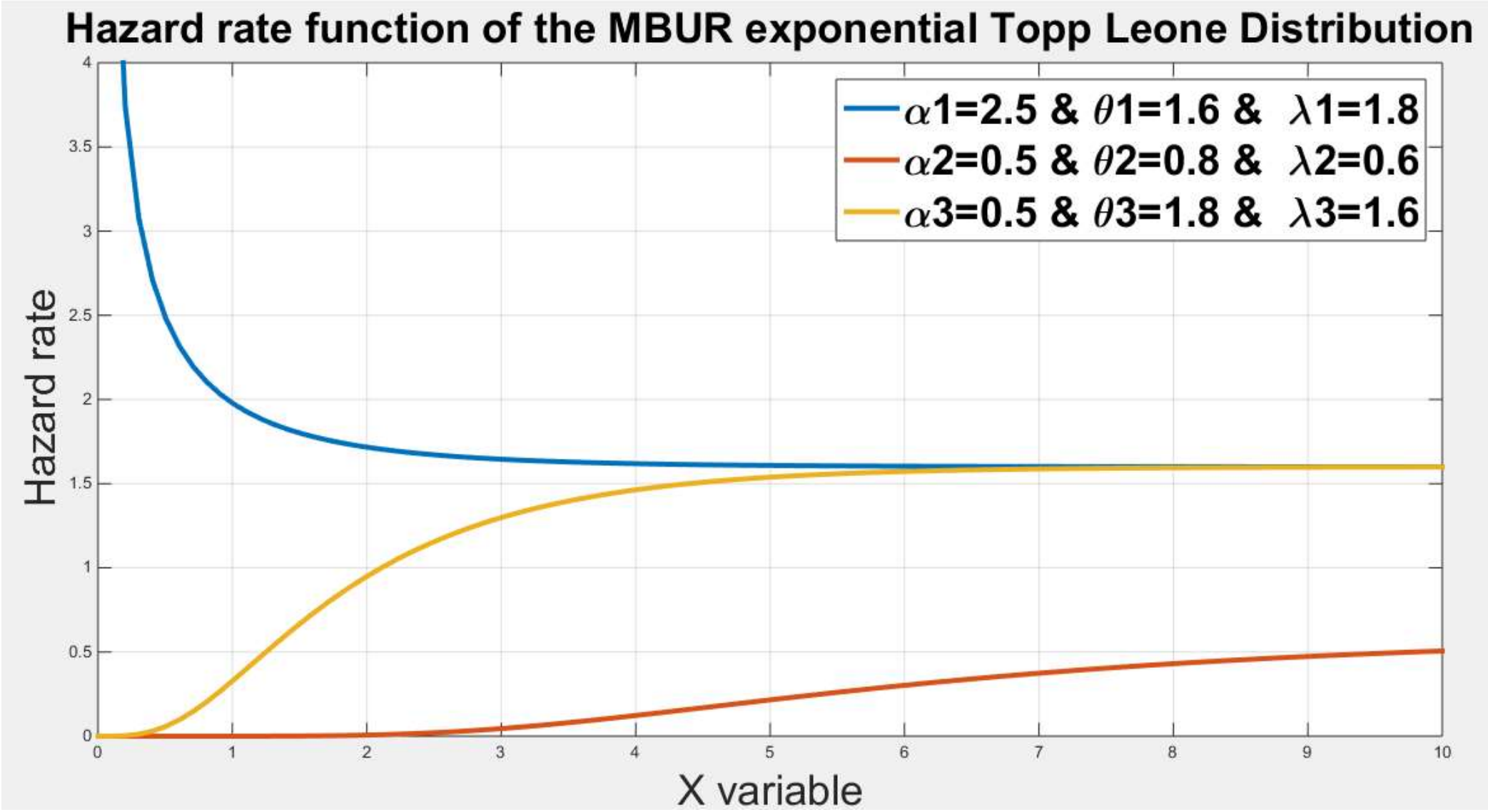

Fig. 18 shows the hazard rate function of the MBUR exponential {Topp Leone} distribution

# Section 4: Real Data Analysis

The dataset is sourced from the OECD, or Organization for Economic Co-operation and Development. It provides information on the economy, social events, education, health, labor, and the environment in the member countries. https://stats.oecd.org/index.aspx?DataSetCode=BLI . The Housing expenditure dataset measures the proportion of household income spent on housing costs. This typically includes rent or mortgage payments, property taxes, and utility bills. It is calculated as the total housing costs divided by the household's gross or net income, expressed as percentage. The real observations were divided by 100 to change them into proportions.

Table 1: The real observations of the Housing expenditure dataset.

| | | | | | | |
|---|---|---|---|---|---|---|
| Australia | Austria | Belgium | Canada | Chile | Costa rica | Czechia |
| 19.4 | 20.8 | 20 | 22.9 | 18.4 | 17 | 23.4 |
| Denmark | Estonia | Finland | France | Germany | Greece | Hungary |
| 23.3 | 17 | 23.1 | 20.7 | 20 | 21.8 | 19.9 |
| Ireland | Italy | Japan | Korea | Latvia | Lithuania | Luxembourg |
| 20.6 | 22.5 | 21.8 | 14.7 | 20.8 | 18.4 | 20.7 |
| Mexico | Nether-land | New Zealand | Norway | Poland | Portugal | Slovak Republic |
| 17.8 | 19.6 | 25 | 17.7 | 21.2 | 19.6 | 27.4 |
| Slovenia | Spain | Sweden | Switzerland | Turkiye | United Kingdom | United States |
| 18.2 | 21.7 | 20.1 | 21.4 | 18.9 | 23.2 | 18.3 |

## 4.1. Descriptive statistics

The mean of the observations is 0.2049, the minimum is 0.1470, the maximum is 0.2740, the standard deviation is 0.0250, the skewness is 0.3278, the kurtosis is 3.8456, the 25$^{th}$ quantile is 0.1852, the median is 0.2060, the 75$^{th}$ quantile is 0.2180. Figure (19) shows the box plot of the dataset with mild right skewness and one outlier in the right upper tail.

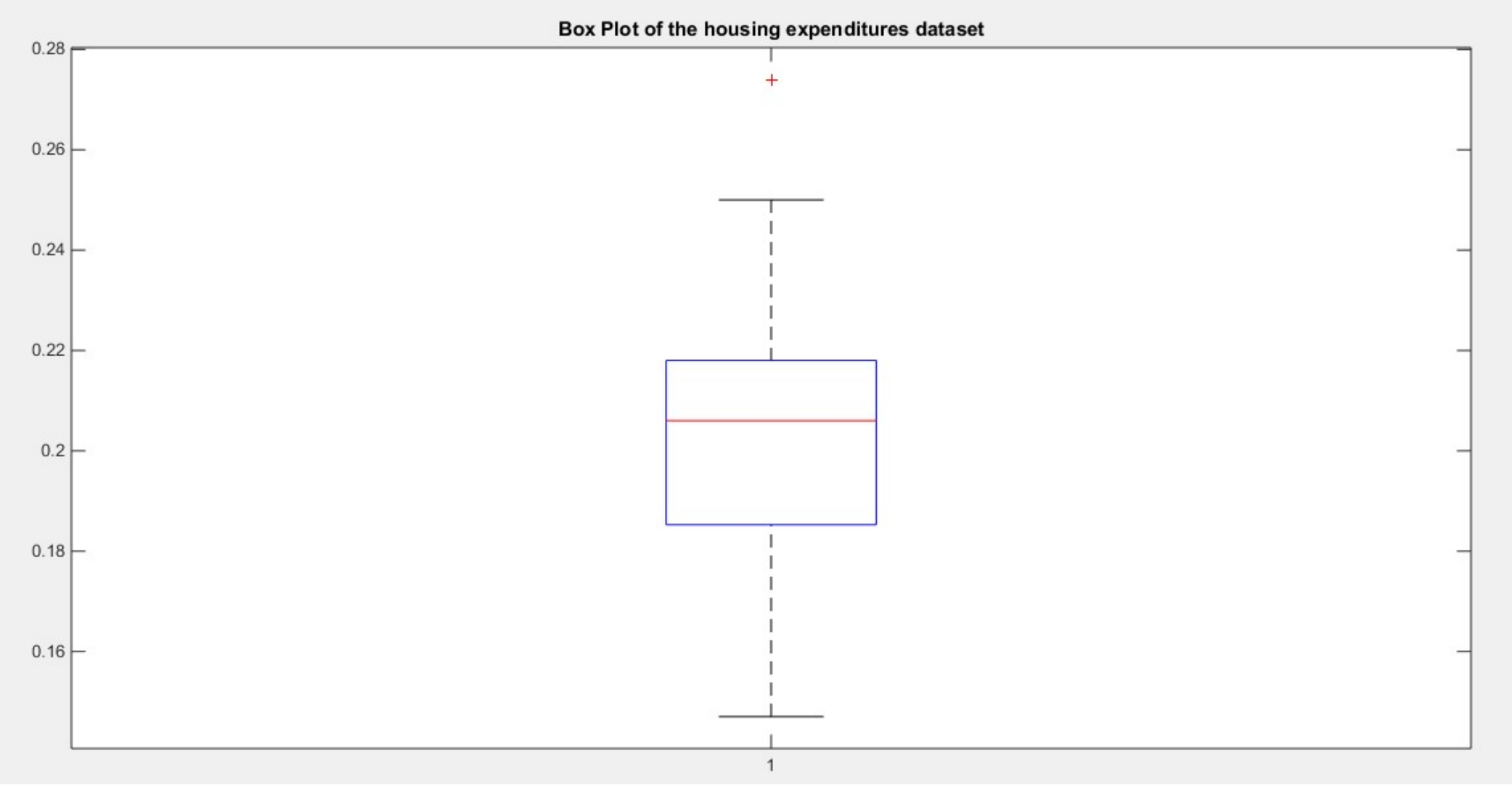

Fig. 19 shows Box Plot for the housing expenditure dataset with one outlier in the upper right tail.

## 4.2. Distributional Data fitting

Table (2): results of housing expenditure dataset fitted by MBUR generated exponential

| Estimate and standard error | | LL | AIC | CAIC | BIC | HQIC | KS | P-value | $H_0$ |
|---|---|---|---|---|---|---|---|---|---|
| $\hat{\alpha}$ | $\hat{\lambda}$ | 79.5272 | -151.0544 | -149.7211 | -144.833 | -148.9068 | 0.0837 | 0.95 | Fail to reject |
| 0.0561 | 30.2955 | | | | | | | | |
| 0.0143 | 2.6264 | | | | | | | | |

$$var - cov\ matrix = \begin{bmatrix} 0.0002 & -0.0365 \\ -0.0365 & 6.8981 \end{bmatrix}$$

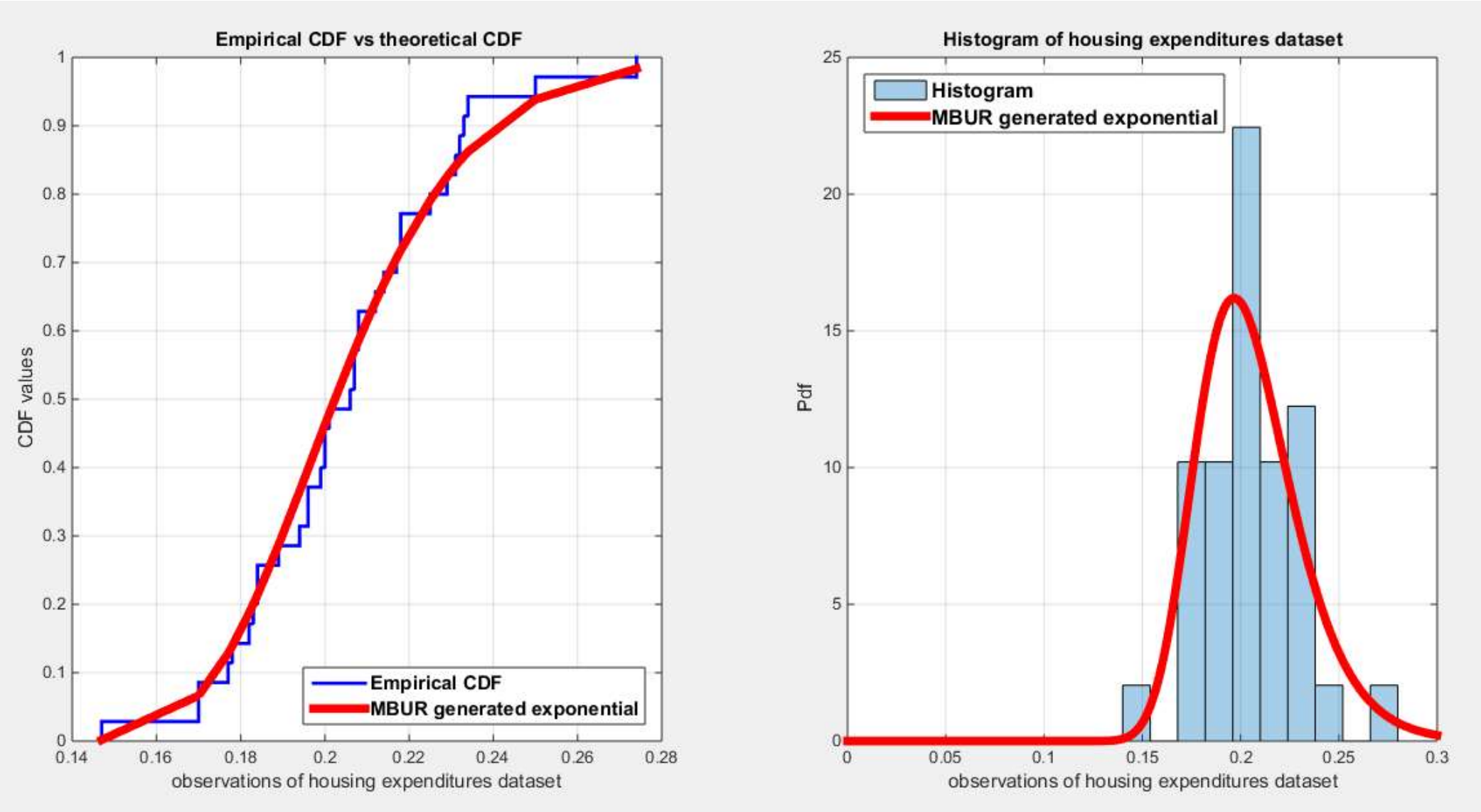

Fig. 20 shows the fitted PDF and CDF for the data. The MBUR generated exponential distribution fits the data well.

Table 2 shows that the MBUR generated exponential distribution fit the data well. This is obvious because the KS test passed the GoF test with p value of 0.95. The LL is 79.5272. Figure 20 shows the fitted CDF and the fitted PDF. Table 3 shows that the MBUR exponential {Kumaraswamy} and the MBUR exponential {Topp Leone} fit the data well. The LL is nearly equal to that gained from fitting the MBUR generated exponential. The values of the AIC, CAIC, BIC, and HQIC are also approximately equal with small differences due to the penalty applied by the numbers of the parameters. As obviously seen these negative values are slightly higher after fitting the MBUR generated exponential than those obtained after fitting the MBUR exponential {Kumaraswamy}, the first distribution has 2 parameters while the later has 4 parameters. The LL attained after fitting the MBUR exponential {Topp Leone} is slightly the highest amongst all the values and this distribution has 3 parameters. And because it is a trade-off between the value of the LL and the number of parameters; the AIC, CAIC, BIC, and HQIC after fitting the MBUR exponential {Topp Leone} are the highest among all the values obtained after fitting the 3 distributions.

Table (3): results of the analysis of the housing expenditure dataset.

| | MBUR-Exponential {Kumaraswamy} | | | | MBUR-Exponential {Topp Leone} | | |
|---|---|---|---|---|---|---|---|
| theta | $\hat{\alpha} = 1.2067e - 4\,, \ SE = 1.4e - 4$ | | | | $\hat{\alpha} = 0.0441, \ SE = 0.0029$ | | |
| | $\hat{\theta} = 2.5883e + 6\,, \ SE = 6e + 6$ | | | | $\hat{\theta} = 60.6068\,, \ SE = 0.6648$ | | |
| | $\hat{a} = 1.9100e + 5\,, SE = 4e + 5$ | | | | $\hat{\lambda} = 2.5985\,, \ SE = 0.0486$ | | |
| | $\hat{b} = 8.3643e + 4\,, SE = 2e + 5$ | | | | | | |
| Variance | 2e-8 | -858 | -63 | -27.7 | 8.4e-6 | 6.28e-4 | 3.2e-5 |
| | -858 | 4e+13 | 3e+12 | 1e+12 | 6.28e-4 | 0.4419 | -0.0130 |
| | -63 | 3e+12 | 2e+11 | 9e+10 | 3.2e-5 | -0.0130 | 0.0024 |
| | -27.7 | 1e+12 | 9e+10 | 4e+10 | | | |
| AIC | -151.0239 | | | | -153.0620 | | |
| CAIC | -149.6906 | | | | -152.2878 | | |
| BIC | -144.8025 | | | | -148.3959 | | |
| HQIC | -148.8763 | | | | -151.4512 | | |
| LL | 79.5120 | | | | 79.5310 | | |
| K-S Value | 0.0864 | | | | 0.0835 | | |
| $H_0$ | Fail to reject | | | | Fail to reject | | |
| P-value | 0.9360 | | | | 0.9506 | | |

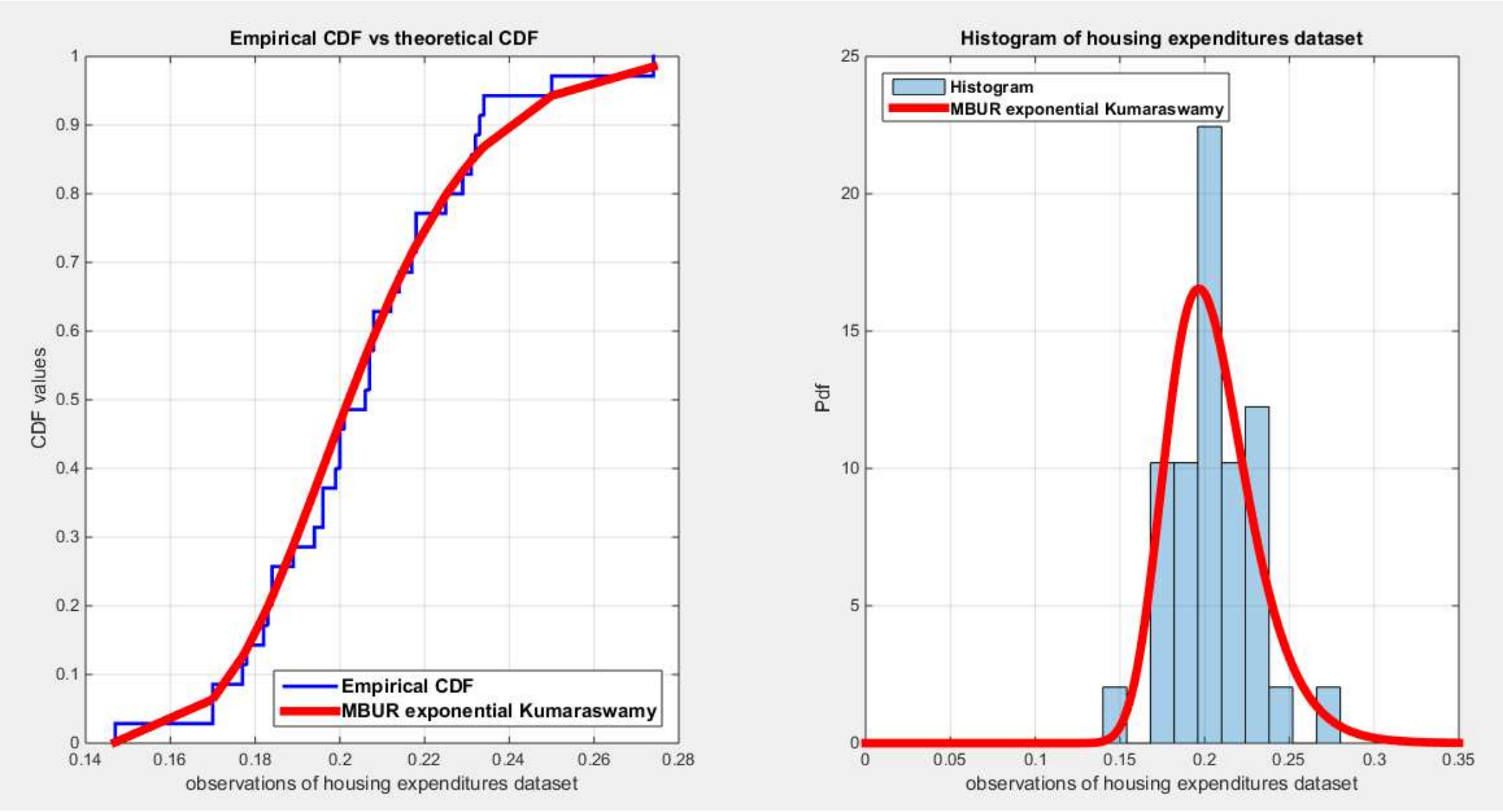

Fig. 21 shows the fitted PDF and fitted CDF for the housing expenditure dataset. MBUR exponential Kumaraswamy fits the data well.

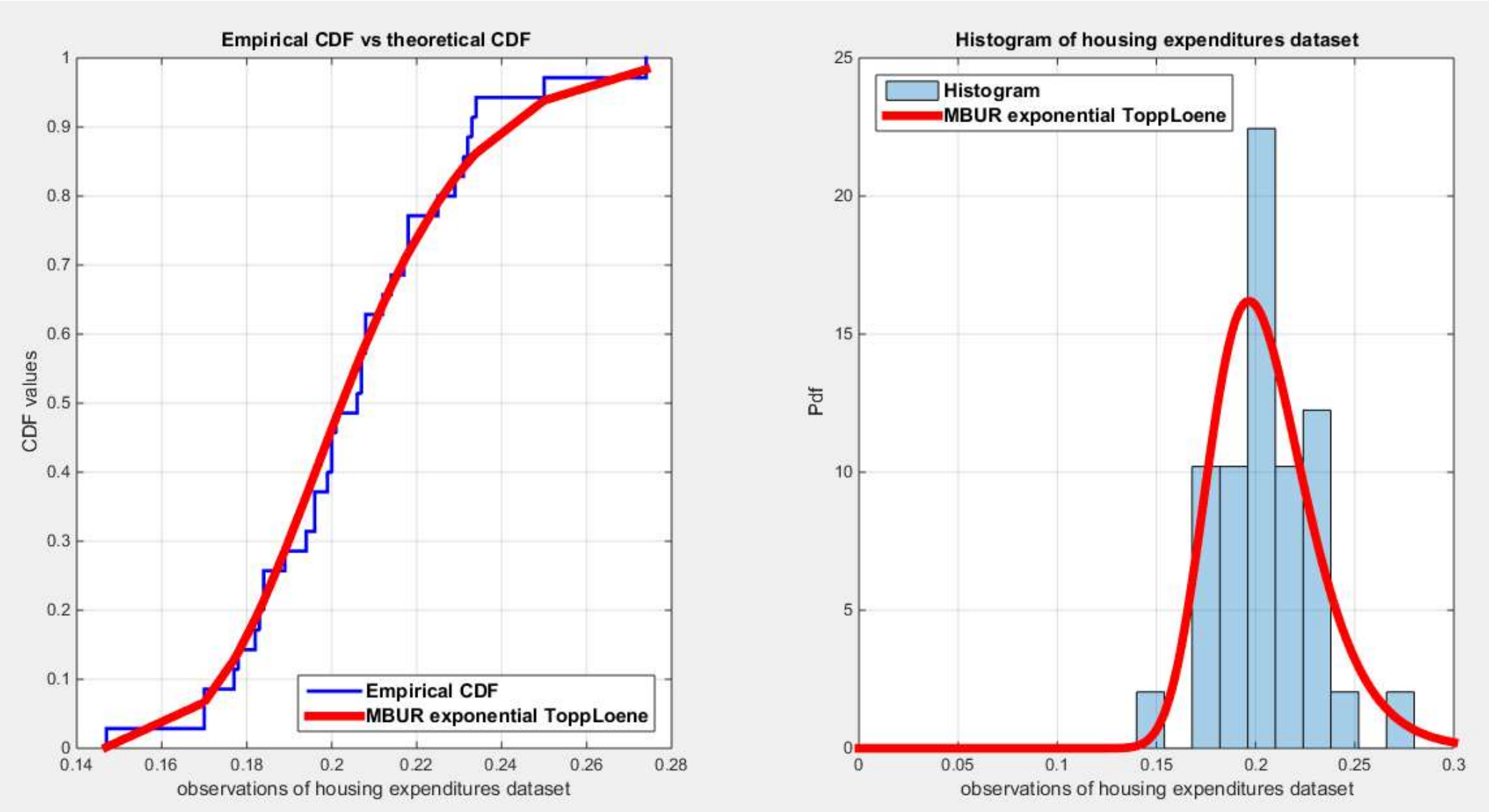

Fig. 22 shows the fitted PDF and fitted CDF for the housing expenditure dataset. MBUR exponential Topp Leone fits the data well.

Figure (21-22) show the fitted CDFs and PDFs for the MBUR exponential {Kumaraswamy} and the MBUR exponential {Topp Leone}. They have nearly identical appearance as the fitted CDF and fitted PDF gained from fitting the MBUR generated exponential distribution.

# Section 5: Conclusion and Future Work

In this paper, the author implanted more than one new parameter in the exponential distribution using the Median Based Unit Rayleigh distribution (MBUR) via different techniques like the generated family concepts and the T-X{Y} procedures. The impact of that addition was noticeably seen in the hazard rate function shape. It is no more constant all through its support. The MBUR generated exponential distribution and the MBUR exponential {Topp Leone} distribution can have initially increasing or decreasing shape followed by a constant rate. While the MBUR exponential {Kumaraswamy} has an initial increasing shape followed by a constant rate resembling the S-shaped appearance. And this differs from the constant hazard rate evident from the exponential distribution. The new distributions inherent the initial increasing or decreasing parts from the MBUR by means of the generated family concept and also from the link quantile distributions, the Kumaraswamy and the Topp Leone, passing through the T-X{Y} procedure. The new distributions can accommodate more data shapes. Hence they can improve the fitness of the data.

To make the picture more advanced, another method for supplementing new parameters like applying the Marshall Olkin generator to any of these new distributions can be considered for future work. Inserting the Marshall Olkin parameter defined on the interval $(0, \infty)$ can be deployed to any of the new distributions. This can be achieved by consuming the CDF and the survival functions from any of the MBUR generated exponential distribution, MBUR exponential {Kumaraswamy}, and the MBUR exponential {Topp Leone} into the Marhall-Olkin generator.

**Declarations:**

**Ethics approval and consent to participate**

Not applicable.

**Consent for publication**

Not applicable

**Availability of data and material**

Not applicable. Data sharing not applicable to this article as no datasets were generated or analyzed during the current study.

**Competing interests**

The author declares no competing interests of any type.

**Funding**

No funding resource. No funding roles in the design of the study and collection, analysis, and interpretation of data and in writing the manuscript are declared

**Authors' contribution**

AI (Attia Iman) carried the conceptualization by formulating the goals, aims of the research article, formal analysis by applying the statistical, mathematical and computational techniques to synthesize and analyze the hypothetical data, carried the methodology by creating the model, software programming and implementation, supervision, writing, drafting, editing, preparation, and creation of the presenting work.

**Acknowledgement**

Not applicable